\documentclass[11pt]{article}  

\usepackage{nicefrac}
\usepackage{bm}
\usepackage{jcappub}
\allowdisplaybreaks

\DeclareMathOperator{\csch}{csch}

\newcommand\kk{{\vphantom{k}}^{\scriptscriptstyle{(3)}}\!k}

\newcommand\Diag{\rm Diag}

\begin{document}


\title{Using Gauge Covariant Lie Derivatives in Poincar\'{e} Gauge and Metric Teleparallel Theories of Gravity}

\author[a]{R. J.  {van den Hoogen}}
\author[a,b,1]{H.  {Forance}}
\note{Alternative email address {x2021cij@stfx.ca}}
\author[a,c]{L. {Taylor}}
\author[a]{M. {Lawton}}

\affiliation[a]{Department of Mathematics and Statistics, St. Francis Xavier University, Antigonish, NS, Canada, B2G 2W5}
\affiliation[b] {Department of Electrical and Computer Engineering, Dalhousie University 
Halifax, NS, Canada, B3H 4R2}
\affiliation[c]{Department of Physics and Astronomy, University of Western Ontario, London, ON, Canada N6A 3K7}

\emailAdd{rvandenh@stfx.ca}
\emailAdd{hudsonxf@dal.ca}
\emailAdd{ltayl59@uwo.ca}
\emailAdd{x2023dwf@stfx.ca}
\date{\today}


\abstract{
A procedure to determine the initial ansatz for the co-frame and spin connection characterizing a Riemann-Cartan geometry respecting a given group of continuous symmetries is illustrated.  Assuming invariance of the metric under a particular group of symmetries, this gauge covariant Lie derivative when applied to the metric and associated orthonormal co-frame determines the values of an antisymmetric compensating matrix.  The gauge covariant Lie derivative of the spin connection depends on this matrix, which when solved yields the spin connection compatible with the chosen co-frame. The procedure is straightforward and can be employed for any Riemann-Cartan geometry having symmetries including those with a non-trivial isotropy subgroup.  Here we illustrate the procedure with numerous examples, including, spherically symmetric, plane symmetric, locally rotationally symmetric Bianchi type III, G\"{o}del, de Sitter and anti-de Sitter geometries.  Further, we have also solved the zero curvature constraint to obtain the resulting spin connection for the corresponding metric teleparallel geometry having this same symmetry group. We complete this investigation by including the Lorentz transformation that yields the proper frame for some of these metric teleparallel geometries. This investigation provides a kinematical framework useful for the further study of Poincar\'{e} Gauge and metric teleparallel theories of gravity.
}

\arxivnumber{2510.03082}
 
\maketitle


\section{Introduction}

\subsection{Isometries in Riemannian geometries}

Invariance under a group of continuous symmetries is often used to reduce the number of variables in complex mathematical and physical problems. In Riemannian theories of gravity such as Einstein's General theory of Relativity (GR), the application of symmetry is robust and well understood.  Since the metric is the fundamental object of interest, symmetries of the geometry and consequently the physics are defined through isometries of the metric \cite{Stephani:2003tm}. The classic example in GR of using symmetries to obtain a physically relevant model, assumes the geometry is static and spherically symmetric with a vacuum matter source which yields the Schwarzschild black hole solution.

In Riemannian geometries, isometries of the metric tensor $\mathbf{g}$ are encoded in the Killing vectors which are solutions to the Killing equations 
\begin{equation} \label{Killing1}
\mathcal{L}_X \mathbf{g} =0.
\end{equation}  
In Riemannian theories of gravity, such as GR, since all geometrical objects are defined via a Lorentz invariant metric, all the geometrical objects computed from the metric are also Lorentz invariant, including the Levi-Civita space-time connection (computed from partial derivatives of the metric) and its corresponding curvature. Further this space-time connection and corresponding curvature are also invariant under the same group of symmetries as the metric\footnote{The Levi-Civita spin connection which depends on the Levi-Civita space-time connection and the choice of frame may not necessarily be invariant under the same group of symmetries as the metric.}. Since the field equations are constructed from the curvature tensor and the metric, they too are invariant under the same group of symmetries as the metric.

In a large class of alternative theories of gravity, (for example, the Poincar\'{e} gauge theory of gravity and its limiting subcase, metric teleparallel gravity), the metric is not the only object of interest, and the application of symmetries in these alternative theories of gravity requires further consideration.  

For example, how should one define a Minkowski geometry in Riemann-Cartan or metric teleparallel geometries?  From a group-theoretical sense, one should define a Minkowski geometry not in terms of the metric, but in terms of the ten-dimensional group of motions that leaves the geometry invariant.  The assumption of a Minkowski metric in Riemann-Cartan geometry is not sufficient to define a geometry devoid of any structures as the torsion is non-trivial.  One also needs additional conditions on the other geometrical components that determine the geometry, in particular, one needs to impose symmetry conditions on the co-frame and/or spin connection which would constrain the torsion to be trivial.

In Riemann-Cartan geometries which provide the geometrical framework for Poincar\'{e} Gauge theories of gravity including the special subcase of metric teleparallel gravity, isometries of the metric only tell part of the story.    As illustrated in the case of a Minkowski geometry, the symmetries of the other fundamental geometrical objects, co-frame and spin connection, should also be considered.  In particular, one needs to contemplate if isometries of the metric extend to the spin connection \cite{FJChinea_1988,Fonseca-Neto:1992xln,fon2002,Hohmann:2019fvf,Hohmann:2019nat,McNutt:2023nxm,McNutt:2024ylz,Hohmann:2024phz}. One of course can construct rich classes of geometries in which the metric is invariant under one group of symmetries, while the spin connection is invariant under a different group, possibly a subgroup.\footnote{See section \ref{TDS} and \ref{TADS} for examples.}  In Riemann-Cartan geometries one has a non-vanishing torsion tensor $\mathbf{T}$ and therefore one reasonable assumption to extend any symmetry present in the metric to the spin connection is to assume  
\begin{equation} \label{affine1}
\mathcal{L}_X \mathbf{T}=0.
\end{equation}
An affine frame symmetry is defined to be any symmetry that satisfies both \eqref{Killing1} and \eqref{affine1}. In this way, the field equations of any Poincar\'{e} gauge theory of gravity derived from an action dependent on the metric, torsion and curvature tensors will also be respectful of these same affine frame symmetries, assuming the matter content also respects such symmetries. 

However, in Riemann-Cartan geometries the application of the Lie derivative to investigate symmetries in the co-frame and spin-connection is not straightforward.  The co-frame and spin connection are not invariant under local Lorentz transformations; therefore using the non-gauge covariant ordinary Lie derivative of the co-frame to investigate continuous symmetries requires some additional considerations in Riemann-Cartan geometries.

\subsection{Affine symmetries in metric affine geometries} 

The metric affine gauge (MAG) approach provides a fully covariant general framework to investigate broad classes of geometries and theories of gravity \cite{Hehl_McCrea_Mielke_Neeman1995}.  However, even within the MAG approach to theories of gravity, given a set of continuous symmetries, how does one determine the metric, the co-frame and the spin-connection which satisfy the desired symmetry requirements. For example, in the broad class of metric affine geometries, one approach summarized in \cite{Obukhov:2015eqa} is to define a \emph{Generalized Killing Vector} by postulating that in addition to the Lie derivative of the metric being zero, the Lie derivative of the distortion tensor\footnote{The distortion tensor is the difference between the affine connection and the Levi-Civita connection.} is also zero.  These \emph{Generalized Killing Vectors} reduce to affine symmetries in the Riemann-Cartan limit when the non-metricity is zero.   However, given a metric invariant under a group of isometries, there is no clear algorithm presented which determines the appropriate co-frame and spin-connection ansatz needed to start an analysis of some geometrical or physical scenario.  

A robust approach to determine the metric, co-frame and spin connection consistent with spherical symmetry in metric affine geometries has been presented by Hohmann \cite{Hohmann:2019fvf}.  Hohmann and his collaborators have also applied this approach to axial symmetry, cosmological symmetries and maximal symmetries with a particular emphasis on metric teleparallel geometries \cite{Hohmann:2019nat,Bahamonde:2021gfp} and have further generalized these ideas to symmetric teleparallel geometries \cite{Hohmann:2024phz}. However, this approach has not been widely applied to other symmetry groups. A complementary but alternative approach has been developed to determine Riemann-Cartan geometries which are invariant under a given group of affine symmetries. This complementary approach uses the Cartan-Karlhede algorithm to fix the freedom in choosing the initial co-frame ansatz as much as possible up to the linear isotropy subgroup which then allows one to define a class of invariantly defined co-frames.  Within this class of invariantly defined co-frames, the corresponding spin connection invariant under the affine group of symmetries can be determined \cite{Coley:2019zld,Coley:2022qug,McNutt:2023nxm,McNutt:2024ylz}.  The most general spherically symmetric (non-static) co-frame and corresponding spin-connection has been determined for Riemann-Cartan geometries and its specialization to metric teleparallel geometries in \cite{McNutt:2023nxm}. Additional applications of this algorithm to de Sitter geometries and to spatially-homogeneous and isotropic geometries in Riemann-Cartan and metric teleparallel geometries can also be found in \cite{McNutt:2023nxm,Coley:2022qug,Coley:2023dbg}.  However, not everyone is familiar with the Cartan-Karlhede algorithm therefore limiting the broader acceptance and employability of this approach to investigate symmetries in Poincar\'{e} Gauge theories of gravity.
 
In this paper we advocate for the use of a gauge covariant Lie derivative \cite{Hehl_McCrea_Mielke_Neeman1995} to analyze continuous symmetries in Riemann-Cartan geometries, including its specialization to metric teleparallel geometries. Gauge covariant Lie derivatives applicable to spinors first appeared in the work of Lichernozwic and Kosmann \cite{Lichnerowicz:1963,Kosmann:1966,Kosmann:1971ugf} and have been further developed by Vandyck and collaborators \cite{Vandyck:1988ei,Vandyck:1988gc,Hurley:1994cfa}. These ideas have been more fully developed by Ort{\'i}n (see \cite{Ortin:2002qb} for details and further references). Currently, these gauge covariant Lie derivatives are also known in the literature as Lie-Lorentz derivatives \cite{Ortin:2002qb}, or Kosmann derivatives \cite{Jacobson:2015uqa} in recognition of the work of Y. Kosmann. These Lie-Lorentz derivatives have been used to investigate the first law of black-hole mechanics \cite{Elgood:2020svt,Elgood:2020mdx,Elgood:2020nls} and other related applications \cite{Figueroa-OFarrill:1999klq,Prabhu:2015vua,Bandos:2023zbs,Ortin:2024emt}.  We will show that the Lie-Lorentz derivative is straightforward to apply given any assumed group of symmetries and results in the same symmetry conditions on the co-frame and spin connection as determined in \cite{Hohmann:2019fvf} when the non-metricity is zero. In this paper, we will apply the Lie-Lorentz derivative to geometries having spherically symmetry, but more importantly, we show how the Lie-Lorentz derivative can be used to expand our repertoire of symmetric Riemann-Cartan geometries. 

%

In Section \ref{symmetry} we define the Lie-Lorentz derivative and how it can be used to define symmetries in Riemann-Cartan geometries.  In Section \ref{examples} we illustrate how to use the Lie-Lorentz derivative to determine the geometrical framework for the metric, co-frame and spin connection given a particular group of  symmetries in Riemann-Cartan and metric teleparallel geometries.  Examples include spherically symmetric, plane symmetric, locally rotationally symmetric Bianchi type III, G\"{o}del, de Sitter and anti-de Sitter geometries.  We also express the proper frame for some of these metric teleparallel geometries in Appendix \ref{AppendixA}. Concluding remarks are found in Section \ref{conclusion}.

\subsection{Notation and geometrical preliminaries}

Here we assume a $4$-dimensional differentiable manifold $M$ with coordinates $\{x^\mu\}$. We use Greek indices $(\mu,\nu,\dots)$ to represent space-time coordinate indices when necessary.  Latin indices $(a,b,\dots)$, represent frame or tangent-space indices. At each point on the manifold there exists a tangent plane to the manifold. On this tangent plane one can define four linearly independent vectors $e_a$. A co-frame field $h^a$ can be defined uniquely through the relation  $e_b\rfloor h^a = \delta^a_b$ where $\rfloor$ indicates the interior product of a vector and a $p$-form. The wedge $\wedge$ denotes the exterior or wedge product. Lengths and angles are determined by the addition of a symmetric metric field $\mathbf{g}$ on $M$, having components $g_{ab}=\mathbf{g}(e_a,e_b)$. Further, by adding additional structure to the manifold, a covariant differentiation process can be defined.  The addition of an independent linear affine connection $\bm{\omega}$, represented as a matrix-valued one-form $\omega^{a}_{\phantom{a}b}$ or spin connection fulfils this requirement.  The geometrical objects $\{g_{ab}, h^a,\omega^{a}_{\phantom{a}b}\}$ are in general independent. However, additional geometrical or physical assumptions yield constraints on and between $\{g_{ab}, h^a, \omega^{a}_{\phantom{a}b}\}$.

Gravitational physics is generally assumed to be invariant under arbitrary changes of coordinates ({\em General Covariance}).  One of the advantages of using the MAG framework, is that it uses Cartan's differential forms and therefore coordinate independence is built into the framework.  A second principle assumed in many theories of gravity is the invariance of the theory under changes in the frame of reference ({\em Principle of Relativity}) \cite{Ortin:2004ms,Hehl_McCrea_Mielke_Neeman1995}. Therefore, in geometrical theories of gravity based on the ({\em Principle of Relativity}), one can apply a $GL(4,\mathbb{R})$ gauge transformation without loss of generality.  This gauge freedom can be used to choose a gauge so that computations are simplified. A convenient choice of gauge useful in Poincar\'{e} Gauge and Metric Teleparallel theories of gravity is the orthonormal gauge in which the tangent space metric is diagonalized to become $g_{ab}=\Diag[-1,1,1,1]$. This gauge choice does not fix everything, the metric is still invariant under local orthogonal transformations $O(1,3)$, or any restrictions thereof, for example, Lorentz transformations $SO(1,3)$. From here on, we shall choose an orthonormal co-frame having a $SO(1,3)$ (Lorentz) residual gauge freedom.  Since the non-metricity is zero in Riemann-Cartan geometries, the components of the spin connection one-form are antisymmetric when written $\omega_{ab}=-\omega_{ba}$. This also means that we only need to compute six components of the spin connection, the other six are determined via the symmetries.

Round brackets surrounding indices represents symmetrization, while square brackets represents anti-symmetrization. Any underlined index is not included in the symmetrization.
The metric signature is assumed to be $(-,+,+,+)$.
The exterior covariant derivative, denoted by $D$, of a matrix-valued p-form $U^a_{~b}$ with respect to the linear spin connection $\omega^a_{~b}$ is
\begin{equation}
DU^a_{~b}:=dU^a_{~b}+\omega^a_{~c}\wedge U^c_{~b}-\omega^c_{~b}\wedge U^a_{~c},
\end{equation}
where $d$ indicates the exterior derivative. 
We shall use $\ell$ to represent the Lie derivative of scalar valued $p$-forms, $\mathcal{L}$ to represent the ordinary Lie derivative, and introduce $\mathcal{K}$ as the gauge covariant Lie-Lorentz or Kosmann derivative.
The Lie derivative of a scalar-valued $p$-form $\pi$ can be computed simply as the anti-commutator of the exterior derivative and the interior product
\begin{equation}
\ell_X \pi = d(X\rfloor \pi) +X\rfloor d\pi.
\end{equation}
For any tensor valued $p$-form $\boldsymbol \Pi$, for example, ${\boldsymbol{\Pi}}  = \Pi^a_{~b} \,e_a\otimes h^b$ the ordinary Lie derivative of $\boldsymbol \Pi$ has components
\begin{equation}
\mathcal{L}_X \Pi^a_{~b} = \ell_X \Pi^a_{~b}-\left(e_c\rfloor \ell_X h^a\right)\Pi^c_{~b} + \left(e_b\rfloor \ell_X h^c\right)\Pi^a_{~c}.
\end{equation}
Since $h^a$ and $\omega^a_{~b}$ are scalar-valued one-forms, we have $\mathcal{L}_X h^a=\ell_X h^a $ and $\mathcal{L}_X \omega^a_{~b}=\ell_X \omega^a_{~b} $ respectively. 


\section{Symmetries in Riemann-Cartan Geometry} \label{symmetry}

\subsection{Challenges with the Lie derivative}

The Lie derivative is a powerful tool in the analysis of gravitational systems with continuous symmetries.  The Lie derivative of a tensorial object $\boldsymbol \Psi$ with respect to some vector field $X$ is simply defined as the rate of change of the pullback of that object along the flow of the vector field 
\begin{equation}
\mathcal{L}_X  \Psi =\lim_{t \to 0} \frac{\phi^*_t{\boldsymbol{\Psi}}- {\boldsymbol{\Psi}}}{t},
\end{equation}
where $\phi^*_t$ is the one parameter local transformation generated by $X$ \cite{Stephani:2003tm}.  One of the advantages of the Lie derivative is that it exists without any additional structure.  If the metric is assumed to be invariant under a continuous symmetry that is generated by some vector field $X$, then $X$ is said to generate an isometry of the metric.  In this case the Lie derivative of the metric along that vector field $X$ is identically zero,
\begin{equation}
\mathcal{L}_X g_{ab}  = 0. 
\end{equation}
In theories of gravity based solely on the metric, i.e., Riemannian theories such as GR, using the Lie derivative to investigate continuous symmetries is not problematic. The metric and its curvature remain invariant under a local Lorentz transformation. The isometries of the metric tell us everything about the geometry.

Since the metric is invariant under the flow of a Killing vector field, a basis of orthonormal co-frames will also remain orthonormal under this same flow. However, individual members of the co-frame basis may undergo a local Lorentz transformation under this flow. That is, the Lie derivative in the direction of the Killing vector $X$ of the orthonormal co-frame basis $h^a$ is not necessarily zero and can be expressed as
\begin{equation}
\mathcal{L}_X h^a = -[\lambda_X]^a_{~b} h^b,
\end{equation}
where the matrix  $[\lambda_X]^a_{~b}$ represents a local Lorentz transformation that depends on the Killing vector $X$. 
The Lie derivative of the co-frame is sensitive to any residual gauge transformations that might exist. Using Lie derivatives to investigate continuous symmetries of objects that transform covariantly under a gauge transformation in Riemann-Cartan geometries requires additional consideration. Further, in the event that the connection is independent of the metric, it becomes necessary to place a constraint on the Lie derivative of the connection typically defining an affine symmetry in the process. 

\subsection{Introducing the Lie-Lorentz derivative}

Given the challenges with using a non-Lorentz covariant Lie derivative to define continuous symmetries in Riemann-Cartan geometries we advocate for a modified approach.  A gauge covariant Lie-Lorentz derivative $\mathcal{K}$ can be defined as follows \cite{Ortin:2002qb,Jacobson:2015uqa}
\begin{equation} \label{Kosman}
\mathcal{K}_X h^a:= \mathcal{L}_X h^a +[\lambda_X]^a_{~b}h^b, 
\end{equation}
where the $[\lambda_X]^a_{~b}$ compensates for the local Lorentz gauge transformation. This Lie-Lorentz derivative of the co-frame transforms covariantly 
\begin{equation}
\mathcal{K}_X h'^a=\Lambda^a_{~b}\left(\mathcal{K}_X h^b\right),
\end{equation}
under a local Lorentz transformation $h^a\to h'^a=\Lambda^a_{~b}h^b$ provided the compensating term $[\lambda_X]^a_{~b}$ transforms as
\begin{equation}\label{transformation_of_lambda}
[\lambda_X]^a_{~b} \to [\lambda_X']^a_{~b}=\Lambda^a_{~c}[\lambda_X]^c_{~d}\Lambda^{-1}{}^d_{~b}+\Lambda^a_{~c}\mathcal{L}_X\Lambda^{-1}{}^c_{~b},
\end{equation}
under a local Lorentz transformation $\Lambda^a_{~b}$. 

The Lie-Lorentz derivative is extended to any Lorentz tensor by assuming that the usual product rule of differentiation extends to the Lie-Lorentz derivative. The Lie-Lorentz derivative of the spin connection one-form is defined so that it compensates for changes in the connection as a result of an infinitesimal Lorentz transformation  
 \cite{Jacobson:2015uqa}
\begin{equation}\label{def_Komega}
\mathcal{K}_X \omega^a_{~b}:=\mathcal{L}_X\omega^a_{~b} - D[\lambda_X]^a_{~b}.
\end{equation}
Again, this Lie-Lorentz derivative of the spin connection transforms covariantly under a Lorentz transformation $\Lambda^a_{~b}$.  If $\omega^a_{~b}\to\omega'^a_{~b}=\Lambda^a_{~c}\omega^c_{~d}\Lambda^{-1}{}^d_{~b}+\Lambda^a_{~c}d\Lambda^{-1}{}^c_{~b}$ under a Lorentz transformation, and given equation \eqref{transformation_of_lambda}, then it can be shown that 
\begin{equation}
\mathcal{K}_X \omega'^a_{~b}=\Lambda^a_{~c} \left(\mathcal{K}_X \omega^c_{~d}\right)\Lambda^{-1}{}^d_{~b}.
\end{equation}

\subsection{Defining continuous symmetries in Riemann-Cartan geometries}

Imposing that the Lie-Lorentz derivative of the metric with respect to some vector field $X$ is zero, 
\begin{equation}\label{Kg}
\mathcal{K}_X g_{ab}:=\mathcal{L}_X g_{ab}-[\lambda_X]^c_{~a}g_{cb}-[\lambda_X]^c_{~b}g_{ac}=0,
\end{equation}
then the symmetric part of the compensating matrix $[\lambda_X]_{ab}$ can be determined 
\begin{equation}
[\lambda_X]_{(ab)}=\frac{1}{2}\mathcal{L}_X g_{ab},
\end{equation}
from the ordinary Lie derivative of the metric. Further, if $X$ generates an isometry of the metric, then  
\begin{equation}
[\lambda_X]_{(ab)}=0.
\end{equation}

To complete the determination of the compensating matrix $[\lambda_X]_{ab}$, choose any orthonormal co-frame basis $h^a$ and assume that the Lie-Lorentz derivative of this co-frame basis with respect to $X$ is also zero,  
\begin{equation}\label{Kh}
\mathcal{K}_X h^a=0.
\end{equation}
The antisymmetric part of the compensating matrix $[\lambda_X]_{ab}$ can be determined via equation \eqref{Kosman} to yield 
\begin{equation} \label{definition of lambda}
[\lambda_X]_{[ab]}=e_{[a}\rfloor {\mathcal{L}}_X h_{b]},
\end{equation}
where $h_a=g_{ab}h^b$.  So given a vector field $X$ generating some isometry of the metric, (i.e., a Killing vector), and selecting an orthonormal co-frame basis $h^a$,  the compensating matrix $[\lambda_X]_{ab}$ can be completely determined by setting both the Lie-Lorentz derivative of the metric and co-frame basis to zero. It is interesting to note that different choices for the orthonormal co-frame basis for a given metric will yield different, but related by transformation \eqref{transformation_of_lambda}, compensating matrices $[\lambda_X]_{ab}$. 

Now in some theories of gravity, the non-Levi-Civita spin connection, $\omega^a_{~b}$, also plays a non-trivial role.  Therefore, to have a completely Lie-Lorentz invariant geometry which includes a non-Levi-Civita spin connection, we require that 
\begin{equation}\label{Lie spin connection}
\mathcal{K}_X \omega^a_{~b}=0,
\end{equation}
which yields a set of differential equations via equation \eqref{def_Komega} that determine the corresponding spin connection associated with the given metric and chosen co-frame basis.  We note, if $X$ is a Killing vector of the metric, then the conditions \eqref{Kg}, \eqref{Kh} and \eqref{Lie spin connection} yield equivalent symmetry conditions employed by Hohmann \cite{Hohmann:2019fvf} in the analysis of symmetries in metric affine geometries when the non-metricity is trivial.

There is a special case worth noting.  Consider a one-dimensional group of symmetries having an isotropy (or any symmetry group where this one-dimensional isotropy group is an element of the center of the symmetry group). For example, an axi-symmetric geometry invariant under rotations about a time-like axis.  In this case, it is possible to choose an initial ansatz for the co-frame in which the compensating matrix is identically zero.  This particular choice just means that the isotropy is encoded within the spin connection and not the co-frame. Alternatively, by applying the necessary Lorentz transformation associated with the isotropy to the initial co-frame ansatz, the isotropy will be made explicit in the transformed co-frame via the non-trivial compensating matrix (or matrices).  

We illustrate the above procedure in a number of cases of interest where the group of symmetries has a non-trivial isotropy group where the isotropy is explicitly observed in the Lie derivative of the co-frame. 


\section{Examples}\label{examples}

\subsection{\texorpdfstring{$G_3$}{G3} on 2-D spatial hyper-surface --- spherically symmetric}\label{Section-SS}

A four dimensional geometry is said to be spherically symmetric if it admits a $G_3(IX)$ of motions acting on two-dimensional space-like surfaces \cite{Stephani:2003tm} and if the non-metric fields inherit the same symmetry. 
Consequently, it has a one dimensional linear isotropy subgroup consisting of spatial rotations.  Assuming the usual spherical coordinates $x^\mu=[t,r,\theta,\phi]$, the three symmetry vector generators  are
\begin{subequations}\label{SS_generators}
\begin{eqnarray}
X_1&=&\sin(\phi)\partial_\theta + \cos(\phi)\cot(\theta)\partial_\phi, \\
X_2&=&-\cos(\phi)\partial_\theta + \sin(\phi)\cot(\theta)\partial_\phi, \\
X_3 &=& \partial_\phi.
\end{eqnarray}
\end{subequations}
The spherically symmetric line element can be expressed without loss of generality as
$$ds^2=g_{\mu\nu}dx^\mu dx^\nu=-A_1(t,r)^2dt^2+A_2(t,r)^2dr^2+A_3(t,r)^2d\theta^2+A_3(t,r)^2\sin^2(\theta)d\phi^2.$$
Choosing to work in the diagonal orthonormal gauge we choose the following co-frame basis
\begin{equation}\label{SSdiagframe}
h^a=\begin{bmatrix}
A_1(t,r)\, dt \\
A_2(t,r)\, dr \\
A_3(t,r)\, d\theta \\
A_3(t,r) \sin(\theta)\, d\phi \end{bmatrix}.
\end{equation}
We label this co-frame choice as the ``diagonal'' orthonormal co-frame since the corresponding tetrad matrix $h^a_{\ \mu}=\Diag[A_1(t,r),A_2(t,r),A_3(t,r),A_3(t,r)\sin(\theta)]$ is diagonal.

The vectors \eqref{SS_generators} are Killing vectors for the metric, $\mathcal{L}_{X_i} g_{ab}=0,\  i\in \{1,2,3\}$. However, in geometries in which the metric is not the primary or sole object describing the geometry, the investigation of continuous symmetries is not as straightforward as it is in Riemannian geometries. For example, in teleparallel geometries, where the co-frame is elevated to be the primary object of interest we observe
\begin{equation}
\mathcal{L}_{X_1} h^a=\begin{bmatrix}
0 \\
0 \\
\hphantom{-}\nicefrac{\cos(\phi)}{\sin(\theta)} h^4   \\
-\nicefrac{\cos(\phi)}{\sin(\theta)} h^3 \end{bmatrix}
,\qquad
\mathcal{L}_{X_2} h^a=\begin{bmatrix}
0 \\
0 \\
\hphantom{-}\nicefrac{\sin(\phi)}{\sin(\theta)} h^4   \\
-\nicefrac{\sin(\phi)}{\sin(\theta)} h^3 \end{bmatrix}
,\qquad 
\mathcal{L}_{X_3} h^a=\begin{bmatrix}
\ 0 \ \\
0 \\
0 \\
0 \end{bmatrix},
\end{equation}
showing clearly how the Lie derivative of the co-frame basis with respect to the Killing vectors $X_1$ and $X_2$ yield a Lorentz transformed co-frame. This observation complicates the determination of the appropriate co-frame ansatz to be used when investigating spherically symmetric geometries in non-Riemannian theories of gravity.

Here we shall assume that continuous symmetries in Riemann-Cartan geometries are defined through the assumption of invariance of the geometrical objects under the Lie-Lorentz derivative,
\begin{equation*}
\mathcal{K}_X g_{ab}=0,\qquad   \mathcal{K}_X h^a=0,\qquad \mbox{and}\qquad \mathcal{K}_X \omega^a_{~b}=0.
\end{equation*}
In the spherically symmetric example considered here, we determine the compensating matrices $[\lambda_{X_i}]^a_{~b},\  i\in\{1,2,3\}$ to be
\begin{subequations}\label{lambda_SS}
\begin{align}
[\lambda_{X_1}]^a_{~b} &=\begin{bmatrix}
\phantom{\quad }0 & \phantom{\quad }0 & 0 & 0\\
\phantom{\quad }0 & \phantom{\quad }0 & 0 & 0\\
\phantom{\quad }0 & \phantom{\quad }0 & 0 & -\nicefrac{\cos(\phi)}{\sin(\theta)}\\
\phantom{\quad }0 & \phantom{\quad }0 & \nicefrac{\cos(\phi)}{\sin(\theta)} & 0 
\end{bmatrix}
,\\
[\lambda_{X_2}]^a_{~b} &=\begin{bmatrix}
\phantom{\quad}0 & \phantom{\quad}0 & 0 & 0\\
\phantom{\quad }0 & \phantom{\quad }0 & 0 & 0\\
\phantom{\quad }0 & \phantom{\quad }0 & 0 & -\nicefrac{\sin(\phi)}{\sin(\theta)}\\
\phantom{\quad }0 & \phantom{\quad }0 & \nicefrac{\sin(\phi)}{\sin(\theta)} & 0 
\end{bmatrix}
,
\end{align}
\end{subequations}
where $[\lambda_{X_3}]^a_{~b}$ only has zero entries.  Since we now know the $[\lambda_X]^a_{~b}$ associated with each Killing vector it then becomes possible to determine the corresponding spin connection that satisfies equation \eqref{Lie spin connection}.

As an example, we compute $\mathcal{K}_{X_i}\omega^3_{~4}=0,\  i\in\{1,2,3\}$.
The equation $\mathcal{K}_{X_3}\omega^3_{~4}=0$ yields
\begin{eqnarray*}
\omega^3_{~4t,\phi} &=& 0, \\
\omega^3_{~4r,\phi} &=& 0, \\
\omega^3_{~4\theta,\phi} &=& 0, \\
\omega^3_{~4\phi,\phi} &=& 0. 
\end{eqnarray*}
The linear combination  $$\sin(\phi)\,\mathcal{K}_{X_1}\omega^3_{~4}-\cos(\phi)\,\mathcal{K}_{X_2}\omega^3_{~4}=0,$$ yields
\begin{eqnarray*}
\omega^3_{~4t,\theta} &=& 0, \\
\omega^3_{~4r,\theta} &=& 0, \\
\omega^3_{~4\theta,\theta} &=& 0, \\
\omega^3_{~4\phi,\theta} &=& \cot(\theta)\omega^3_{~4\phi}+\csc(\theta).
\end{eqnarray*}
The linear combination  $$ \cos(\phi)\,\mathcal{K}_{X_1}\omega^3_{~4}+\sin(\phi)\,\mathcal{K}_{X_2}\omega^3_{~4}=0,$$ yields 
\begin{eqnarray*}
\omega^3_{~4t,\phi} &=& 0, \\
\omega^3_{~4r,\phi} &=& 0, \\
\omega^3_{~4\theta,\phi} &=&  \csc(\theta)\sec(\theta)\omega^3_{~4\phi}+\csc(\theta),\\
\omega^3_{~4\phi,\phi} &=& -\omega^3_{~4\theta}\tan(\theta).
\end{eqnarray*}
The general solution to the above system of partial differential equations is 
$$\omega^3_{~4t}= W_7(t,r),\ \omega^3_{~4r}= W_8(t,r),\ \omega^3_{~4\theta}=0,\ \omega^3_{~4\phi}=-\cos(\theta),
$$ 
where $W_7(t,r)$ and $W_8(t,r)$ are two arbitrary functions. 

Solving the remaining equations \eqref{Lie spin connection}, the most general spin connection for Riemann-Cartan geometries assuming a diagonal orthonormal gauge respecting spherical symmetry is
given by

\begin{subequations} \label{orthonormal_connection}
\text{\bf Riemann-Cartan} 
\begin{align}
\omega^1_{~2}&=W_1(t,r)\,dt+W_2(t,r)\,dr \\
\omega^1_{~3}&=W_3(t,r)\,d\theta-W_4(t,r)\sin(\theta)\,d\phi \\
\omega^1_{~4}&=W_4(t,r)\,d\theta+W_3(t,r)\sin(\theta)\,d\phi \\
\omega^2_{~3}&=W_5(t,r)\,d\theta-W_6(t,r)\sin(\theta)\,d\phi \\
\omega^2_{~4}&=W_6(t,r)\,d\theta+W_5(t,r)\sin(\theta)\,d\phi \\
\omega^3_{~4}&=W_7(t,r)\,dt+W_8(t,r)\,dr -\cos(\theta)\,d\phi 
\end{align}
\end{subequations}
where $W_j(t,r),\  j \in \{1,2,3,4,5,6,7,8\}$ are eight arbitrary functions.

Restricting ourselves to the metric-teleparallel class of geometries, in which the curvature of the connection is zero, we obtain the most general spin connection for metric teleparallel geometries assuming the diagonal orthonormal gauge respecting spherical symmetry to be

\begin{subequations} \label{orthonormal_connection_flat}
\text{\bf Metric Teleparallel} 
\begin{align}
\omega^1_{~2} &= \hphantom{-} \Big(\partial_{t} \psi(t, r)\Big) \,dt + \Big(\partial_{r} \psi(t, r)\Big) \,dr, \\
\omega^1_{~3} &= -\Big(\sinh{\psi(t, r)} \cos{\chi(t, r)}\Big)\,d\theta -\Big(\sinh{\psi(t, r)} \sin{\chi(t, r)}\sin\theta\Big)\,d\phi, \\
\omega^1_{~4} &= \hphantom{-}\Big(\sinh{\psi(t, r)} \sin{\chi(t, r)}\Big)\,d\theta -(\sinh{\psi(t, r)} \cos{\chi(t, r)}\sin\theta)\,d\phi, \\
\omega^2_{~3} &= \hphantom{-}  \Big(\cosh{\psi(t, r)} \cos{\chi(t, r)}\Big)\,d\theta +\Big(\cosh{\psi(t, r)} \sin{\chi(t, r)}\sin\theta\Big)\,d\phi, \\
\omega^2_{~4} &= -\Big(\cosh{\psi(t, r)} \sin{\chi(t, r)}\Big)\,d\theta +\Big(\cosh{\psi(t, r)} \cos{\chi(t, r)}\sin\theta\Big)\,d\phi, \\
\omega^3_{~4}  &=-\Big(\partial_{t} \chi(t, r)\Big) \,dt - \Big(\partial_{r} \chi(t, r)\Big)\,dr - \Big(\cos{\theta}\Big)\,d\phi
\end{align}
\end{subequations}
where $\psi(t,r)$ and $\chi(t,r)$ are two arbitrary functions.  This result is in complete agreement with the spin connection expressed in \cite{McNutt:2023nxm} which was determined using a different algorithm. The spherically symmetric proper frame for metric-teleparallel geometries in which the spin connection is identically zero has a non-diagonal co-frame and can be found in \cite{vandenHoogen:2024ash}.  Further, it was shown in \cite{vandenHoogen:2024ash} that the proper co-frame associated with the coframe/spin-connection pair given by equations \eqref{SSdiagframe} and \eqref{orthonormal_connection_flat}, which was earlier presented in \cite{McNutt:2023nxm}, is equivalent to the coframe in the Weitzenbock gauge (equation 68 in \cite{Hohmann:2019nat}) upon the application of a remaining coordinate freedom.

\subsubsection{\texorpdfstring{$G_4$}{G4}  --- static spherically symmetric}

This case is also a simple extension of the case of Section \ref{Section-SS} in which in addition to the three
Killing vectors, there exists a fourth Killing vector $X_4 = \partial_t$.
In this case the isotropy group is the same as in the general spherically symmetric case.
Having this fourth Killing vector implies that the three metric functions $A_i(t,r), i\in \{1,2,3\}$ and  the eight arbitrary functions $W_j(t,r),\ j\in \{1,2,3,4,5,6,7,8\}$ are only functions of $r$ in Riemann-Cartan geometries. In the metric teleparallel geometries in addition to the three metric functions being only a function of $r$, the functions $\psi(t,r)$ and $\chi(t,r)$ are also only functions of $r$.

\subsubsection{\texorpdfstring{$G_4$}{G4} on 3-D hyper-surfaces --- Kantowski-Sachs}

This case is also a simple extension of the case of Section \ref{Section-SS} in which in addition to the three
Killing vectors, there exists a fourth Killing vector $X_4 = \partial_r$.
In this case the isotropy group is the same as in the general spherically symmetric case.
Having this fourth Killing vector implies that the three metric functions $A_i(t,r), i\in \{1,2,3\}$ and  the eight arbitrary functions $W_j(t,r),\ j\in \{1,2,3,4,5,6,7,8\}$ are only functions of $t$ in Riemann-Cartan geometries. In the metric teleparallel geometries in addition to the three metric functions being only a function of $t$, the functions $\psi(t,r)$ and $\chi(t,r)$ are also only functions of $t$.

\subsection{\texorpdfstring{$G_3$}{G3} on 2-D spatial hyper-surfaces --- plane symmetric}

A four dimensional geometry is said to be plane symmetric if it admits a $G_3(VII_0)$ of motions acting on two-dimensional space-like surfaces \cite{Stephani:2003tm} and if the non-metric fields inherit the same symmetry.  Consequently, it has a one dimensional linear isotropy subgroup again consisting of spatial rotations.  Choosing Cartesian coordinates $x^\mu=[t,x,y,z]$, the three vector generators for the plane symmetry  are
\begin{subequations}\label{Plane_generators}
\begin{eqnarray}
X_1&=&\partial_y, \\
X_2&=&\partial_z, \\
X_3 &=&z\partial_y-y\partial_z.
\end{eqnarray}
\end{subequations}
The corresponding line-element can be expressed without loss of generality as
\begin{equation}
ds^2 = -A_1(t,x)^2dt^2 + A_2(t,x)^2dx^2+A_3(t,x)^2 (dy^2+dz^2).
\end{equation}
Choosing to work in the diagonal orthonormal gauge we choose the following co-frame basis
\begin{equation}\label{Plane-frame}
h^a=\begin{bmatrix}
A_1(t,x)\, dt \\
A_2(t,x)\, dx \\
A_3(t,x)\, dy \\
A_3(t,x)\, dz \end{bmatrix}.
\end{equation}
 
The vectors \eqref{Plane_generators} are Killing vectors, $\mathcal{L}_{X_i} g_{ab}=0,\  i \in \{1,2,3\}$.  Applying the Lie-Lorentz derivative to the diagonal orthonormal coframe \eqref{Plane-frame} and setting $\mathcal{K}_X h^a=0$ we are able to determine the $[\lambda_{X_i}]^a_{~b},\  i \in \{1,2,3\}$ and from there we can solve $\mathcal{K}_X \omega^a_{~b}=0$ which will determine the corresponding spin connection for our diagonal orthonormal coframe gauge choice.

From equation \eqref{definition of lambda}, the $[\lambda_{X_i}]^a_{~b}$ matrices are
\renewcommand{\arraystretch}{1.2}
\begin{equation}
[\lambda_{X_3}]^a_{~b} =\begin{bmatrix}
0 & 0 & 0 & 0\\
0 & 0 & 0 & 0\\
0 & 0 & 0 & -1\\
0 & 0 & 1 & 0 
\end{bmatrix}
\end{equation}
where $[\lambda_{X_1}]^a_{~b}$ and $[\lambda_{X_2}]^a_{~b}$ only have zero entries. With this information it becomes straight forward to determine the corresponding spin connection via equation \eqref{Lie spin connection}. 
Solving equation \eqref{Lie spin connection} we find the most general spin connection for Riemann-Cartan geometries in the diagonal orthonormal gauge respecting plane symmetry to be

\begin{subequations}
\text{\bf Riemann-Cartan} 
\begin{eqnarray}
\omega^1_{~2}&=W_1(t,x)\,dt+W_2(t,x)\,dx, \\
\omega^1_{~3}&=W_3(t,x)\,dy-W_4(t,x)\,dz, \\
\omega^1_{~4}&=W_4(t,x)\,dy+W_3(t,x)\,dz, \\
\omega^2_{~3}&=W_5(t,x)\,dy-W_6(t,x)\,dz, \\
\omega^2_{~4}&=W_6(t,x)\,dy+W_5(t,x)\,dz, \\
\omega^3_{~4}&=W_7(t,x)\,dt+W_8(t,x)\,dx, 
\end{eqnarray}
\end{subequations}
where $W_j(t,x),\  j \in \{1,2,3,4,5,6,7,8\}$ are eight arbitrary functions.

Additionally, restricting ourselves to the metric-teleparallel class of geometries, in which the curvature of the connection is zero, we obtain a pair of two parameter families of spin connections. For metric teleparallel plane symmetric geometries assuming the diagonal orthonormal gauge the spin connection is

\begin{subequations}
\text{\bf Metric Teleparallel (case 1)}\label{connection-plane} 
\begin{eqnarray}
\omega^1_{~2}&=&-\partial_t(\ln(W(t,x))\, dt -\partial_x(\ln(W(t,x))\,dx\\
\omega^1_{~3}=\omega^2_{~3}&=&W(t,x)\cos(\chi(t,x))\,dy-W(t,x)\sin(\chi(t,x))\,dz \\
\omega^1_{~4}=\omega^2_{~4}&=&W(t,x)\sin(\chi(t,x))\,dy+W(t,x)\cos(\chi(t,x))\,dz \\
\omega^3_{~4}&=&\partial_t(\chi(t,x))\,dt+\partial_x(\chi(t,x))\,dx 
\end{eqnarray}
\end{subequations}
where $W(t,x)$ and $\chi(t,x)$ are two arbitrary functions, or 

\begin{subequations}\label{connection-plane2}
\text{\bf Metric Teleparallel (case 2)} 
\begin{eqnarray}
\omega^1_{~2}&=&\partial_t(\psi(t,x))\,dt+\partial_x(\psi(t,x))\,dx\\
\omega^1_{~3}=\omega^2_{~3}&=&0 \\
\omega^1_{~4}=\omega^2_{~4}&=&0 \\
\omega^3_{~4}&=&\partial_t(\chi(t,x))\,dt+\partial_x(\chi(t,x))\,dx 
\end{eqnarray}
\end{subequations}
where $\psi(t,x)$ and $\chi(t,x)$ are two arbitrary functions. The Lorentz transformation that yields the plane symmetric proper frame for metric-teleparallel geometries in which the spin connection is identically zero can be found in Appendix \ref{appendix-proper-plane}.

\subsection{\texorpdfstring{$G_4$}{G4}  on 3-D hyper-surfaces --- LRS Bianchi III}
The Locally Rotationally Symmetric (LRS) Bianchi type III geometry is an example of a geometry that admits four Killing vectors acting on three dimensional spatial hypersurface.  Other examples of LRS spatially homogeneous geometries exist but are not included here, however the procedure to determine the appropriate frame and spin connection is the same. 

With the coordinate choice $x^\mu=[t,x,y,z]$, the Killing vectors are
\begin{subequations}\label{LRSIII_generators}
\begin{eqnarray}
X_1&=& \partial_y \\
X_2&=&\partial_z, \\
X_3&=&\partial_x - y\partial_y, \\
X_4&=&y\partial_x-\left(\frac{y^{2}}{2}-\frac{e^{-2x}}{2}\right)\partial_y.
\end{eqnarray}
\end{subequations}
The generic line element for LRS Bianchi type III can be written as follows with arbitrary functions $A_1(t)$ and $A_2(t)$
\begin{equation}
ds^2=-dt^2+A_1(t)^2dx^2+A_1(t)^2e^{2x}dy^2+A_2(t)^2dz^2.
\end{equation}
Choosing again to work in the diagonal orthonormal gauge, we define the following co-frame basis
\begin{equation}\label{LRSIII-frame}
h^a=\begin{bmatrix} 
 dt  \\
A_1(t)\, dx \\
A_1(t)e^{x}\, dy \\
A_2(t)  \, dz \end{bmatrix}.
\end{equation}

Applying the Lie-Lorentz derivative to the orthonormal coframe \eqref{LRSIII-frame} and setting $\mathcal{K}_X h^a=0$ we determine the matrices $[\lambda_{X_i}]^a_{~b},\  i\in\{1,2,3,4\}$ and from there we solve $\mathcal{K}_X \omega^a_{~b}=0$ to determine the corresponding spin connection for our orthonormal coframe gauge choice.
From equation \eqref{definition of lambda}, the $[\lambda_{X_i}]^a_{~b}$ matrices are
\renewcommand{\arraystretch}{1.2}
\begin{equation}
[\lambda_{X_4}]^a_{~b} =\begin{bmatrix}
0 & 0 & 0 & 0\\
0 & 0 & -e^{-x} & 0\\
0 & e^{-x} & 0 & 0\\
0 & 0 & 0 & 0 
\end{bmatrix},
\end{equation}
where $[\lambda_{X_i}]^a_{~b},\ i\in\{1,2,3\}$ only have zero entries.

Solving equation \eqref{Lie spin connection} we find the most general spin connection for Riemann-Cartan LRS Bianchi type III geometries in the orthonormal gauge having a $G_4$ group of isometries is

\begin{subequations}\label{RCconnectionIII}
\text{\bf Riemann-Cartan} 
\begin{eqnarray}
\omega^1_{~2}&= &W_1(t)\,dx+W_2(t)e^x\,dy, \\
\omega^1_{~3}&= &-W_2(t)\,dx+ W_1(t)e^{x} \,dy, \\
\omega^1_{~4}&= &W_3(t)\,dt+W_4(t)\,dz, \\
\omega^2_{~3}&= &-W_5(t)\,dt-e^{x}\,dy+W_6(t)\,dz, \\
\omega^2_{~4}&= &W_7(t)\,dx+W_{8}(t)e^{x}\,dy, \\
\omega^3_{~4}&=&-W_{8}(t)\,dx+W_{7}(t)e^{x}\,dy ,
\end{eqnarray}
\end{subequations}
where  $W_j(t),\  j\in\{1,2,3,4,5,6,7,8\}$ are eight arbitrary functions of $t$.

If we further restrict ourselves to the metric-teleparallel class of geometries, in which the curvature of the connection is zero, then we find a spin connection for metric teleparallel geometries assuming the orthonormal gauge for the LRS Bianchi type III geometry to be given by the substitution

\begin{subequations}\label{solutionLRSIII-NULLCURV}
\text{\bf Metric Teleparallel} 
\begin{align}
W_1(t)&=\cos(\chi(t))\cosh(\psi(t)), &   W_2(t)&=\sin(\chi(t))\cosh(\psi(t)),\\
W_3(t)&=\partial_t{\psi}(t),  &   W_4(t)&=0,\\
W_5(t)&=\partial_t{\chi}(t), &   W_6(t)&=0,\\
W_7(t)&=\cos(\chi(t))\sinh(\psi(t)),   &   W_{8}(t)&=\sin(\chi(t))\sinh(\psi(t)).
\end{align}
\end{subequations}
into equation \eqref{RCconnectionIII}.
The arbitrary functions  $\psi(t)$ and $\chi(t)$ act as boost and rotation parameters.  The Lorentz transformation that yields the LRS Bianchi type III proper frame for metric-teleparallel geometries in which the spin connection is identically zero can be found in Appendix \ref{appendix-proper-LRS-Bianchi}.

\subsection{\texorpdfstring{$G_5$}{G5} --- G\"{o}del}

The G\"{o}del geometry is a homogeneous geometry having five Killing vectors acting on the entire four dimensional geometry. Choosing Cartesian coordinates $x^\mu=[t,x,y,z]$, the five vector generators  are
\begin{subequations}\label{Godel_generators}
\begin{eqnarray}
X_1&=&2e^{-x}\partial_t-z\partial_x+\left(\frac{z^2}{2}-e^{-2x}\right)\partial_z, \\
X_2&=&\partial_t, \\
X_3&=&\partial_x-z\partial_z, \\
X_4&=&\partial_y, \\
X_5&=&\partial_z.
\end{eqnarray}
\end{subequations}
The line element \cite{Stephani} can be expressed as
\begin{equation}
ds^2=A^2\left[-(dt+e^xdz)^2+dx^2+dy^2+\frac{1}{2}e^{2x}dz^2\right],
\end{equation}
where $A$ is a constant.
We choose the orthonormal frame
\begin{equation}\label{Godel-frame}
h^a=\begin{bmatrix} 
A\, (dt +e^x\,dz) \\
A\, dx \\
A\, dy \\
\frac{A}{\sqrt{2}}e^x  \, dz \end{bmatrix}.
\end{equation}

Applying the Lie-Lorentz derivative to the orthonormal coframe \eqref{Godel-frame} and setting $\mathcal{K}_X h^a=0$ we determine the $[\lambda_{X_i}]^a_{~b},\  i\in\{1,2,3,4,5\}$ and from there we solve $\mathcal{K}_X \omega^a_{~b}=0$ which determines the corresponding spin connection for our orthonormal coframe gauge choice.
From equation \eqref{definition of lambda}, the $[\lambda_{X_i}]^a_{~b}$ matrices are
\renewcommand{\arraystretch}{1.2}
\begin{equation}
[\lambda_{X_1}]^a_{~b} =\begin{bmatrix}
0 & 0 & 0 & 0\\
0 & 0 & 0 & \sqrt{2}e^{-x}\\
0 & 0 & 0 & 0\\
0 & -\sqrt{2}e^{-x} & 0 & 0 
\end{bmatrix},
\end{equation}
where $[\lambda_{X_i}]^a_{~b},\ i\in\{2,3,4,5\}$ only have zero entries.

Solving equation \eqref{Lie spin connection} we find the most general spin connection for Riemann-Cartan geometries in the orthonormal gauge having a $G_5$ group of isometries

\begin{subequations}\label{spin-connection-godel}
\text{\bf Riemann-Cartan} 
\begin{eqnarray}
\omega^1_{~2}&=&W_1\,dx-\frac{1}{\sqrt{2}}W_2e^x\,dz, \\
\omega^1_{~3}&=&W_3\,(dt+e^x\,dz)+ W_4 \,dy, \\
\omega^1_{~4}&=&W_2\,dx+\frac{1}{\sqrt{2}}W_1e^x\,dz, \\
\omega^2_{~3}&=&W_5\,dx+\frac{1}{\sqrt{2}}W_6e^x\,dz, \\
\omega^2_{~4}&=&W_7\,(dt+e^x\,dz)+W_8\,dy -\frac{1}{\sqrt{2}}e^x\,dz, \\
\omega^3_{~4}&=&W_6\,dx-\frac{1}{\sqrt{2}}W_5e^x\,dz ,
\end{eqnarray}
\end{subequations}
where $W_j,\  j\in\{1,2,3,4,5,6,7,8\}$ are eight arbitrary constants. 
 
If we further restrict ourselves to the metric-teleparallel class of geometries, then we find a spin connection for metric teleparallel G\"{o}del geometries in the orthonormal gauge.    
Solving the null curvature condition for the spin connection, we obtain a pair of two-parameter families of spin connections. For metric teleparallel G\"{o}del geometries assuming the orthonormal gauge the spin connection can be determined through the following substitutions in \eqref{spin-connection-godel}

\begin{subequations}\label{Godel-spin-1}
\text{\bf Metric Teleparallel (case 1)} 
\begin{eqnarray}
W_1&=& W_2=W_3=0,\\
W_5&=& W_6=0,\\
W_7&=&\frac{1}{\sqrt{2}},
\end{eqnarray}
\end{subequations}
where $W_4$ and $W_8$ remain as arbitrary constants, or,

\begin{subequations}\label{Godel-spin-2}
\text{\bf Metric Teleparallel (case 2)} 
\begin{eqnarray}
W_1&=&\cosh(\psi)\cos(\chi),\\
W_2&=&\cosh(\psi)\sin(\chi),\\
W_3&=& W_4=0,\\
W_5&=&\sinh(\psi)\cos(\chi),\\
W_6&=&-\sinh(\psi)\sin(\chi),\\
W_7&=&W_8=0,
\end{eqnarray}
\end{subequations}
where $\psi$ and $\chi$ are arbitrary constants. The Lorentz transformation needed to compute the proper frame corresponding to each of these spin connection choices can be found in Appendix \ref{appendix-proper-godel}.

\subsection{\texorpdfstring{$G_6$}{G6} on 3-D hyper-surfaces --- spatially homogeneous and isotropic} \label{RW}

A four dimensional geometry having a six dimensional group of motions acting on three-dimensional space-like hyper-surfaces is said to be spatially homogenous and isotropic. These geometries are heavily used in building cosmological models. In this case there exists a three dimensional linear isotropy group consisting of spatial rotations.  Choosing coordinates $x^\mu=[t,r,\theta,\phi]$, the six vector generators for the $G_6$ symmetry are given by equation \eqref{SS_generators} plus an additional three killing vectors
\begin{subequations}\label{TRW_generators}
\begin{align}
X_1&=\sin(\phi)\partial_\theta + \cos(\phi)\cot(\theta)\partial_\phi, \\
X_2&=-\cos(\phi)\partial_\theta + \sin(\phi)\cot(\theta)\partial_\phi, \\
X_3&= \partial_\phi\\
X_4&= \sqrt{1-\kk r^2}\sin(\theta)\sin(\phi)\partial_r +\frac{\sqrt{1-\kk r^2}}{r}\cos(\theta)\sin(\phi)\partial_\theta +\frac{\sqrt{1-\kk r^2}}{r}\frac{\cos(\phi)}{\sin(\theta)}\partial_\phi , \\
X_5&= \sqrt{1-\kk r^2}\sin(\theta)\cos(\phi)\partial_r +\frac{\sqrt{1-\kk r^2}}{r}\cos(\theta)\cos(\phi)\partial_\theta -\frac{\sqrt{1-\kk r^2}}{r}\frac{\sin(\phi)}{\sin(\theta)}\partial_\phi , \\
X_6&= \sqrt{1-\kk r^2}\cos(\theta)\partial_r -\frac{\sqrt{1-\kk r^2}}{r}\sin(\theta)\partial_\theta, 
\end{align}
where the discrete parameter $\kk$ can take on values $-1,0,1$ corresponding to the curvature of the three dimensional spatial hyper-surface.
\end{subequations}
The corresponding line-element can be expressed without loss of generality as
\begin{equation}
ds^2 = -dt^2 + \frac{A(t)^2}{1-\kk r^2}\,dr^2+A(t)^2r^2\,d\theta^2+A(t)^2r^2\sin^2(\theta)\,d\phi^2.
\end{equation}
We can choose a diagonal orthonormal gauge for the co-frame basis
\begin{equation} \label{TRW-frame}\renewcommand{\arraystretch}{1.4}
h^a=\begin{bmatrix}
  dt \\
\frac{A(t)}{\sqrt{1-\kk r^2}} \,dr \\
A(t)r \,d\theta \\
A(t)r \sin(\theta)\,d\phi \end{bmatrix}.
\end{equation}

Aplying the Lie-Lorentz derivative to the diagonal orthonormal coframe \eqref{TRW-frame} and setting $\mathcal{K}_X h^a=0$ we determine the $[\lambda_{X_i}]^a_{~b},\ i\in\{1,2,3,4,5,6\}$ and from there we solve $\mathcal{K}_X \omega^a_{~b}=0$ which will determine the corresponding spin connection for our diagonal orthonormal coframe gauge choice.

From equation \eqref{definition of lambda}, the $[\lambda_{X_i}]^a_{~b}$, $i\in \{1,2\}$ are given by equation \eqref{lambda_SS} and $[\lambda_{X_3}]^a_{~b}$ only has zero entries.  The other $[\lambda_{X_i}]^a_{~b}$, $i\in\{4,5,6\}$ matrices are
\begin{subequations}
\renewcommand{\arraystretch}{1.2}
\begin{align}
[\lambda_{X_4}]^a_{~b} &=\begin{bmatrix}
\phantom{\quad}0\phantom{\quad} & \phantom{\quad}0\phantom{\quad} & \phantom{\quad}0\phantom{\quad} & \phantom{\quad}0\phantom{\quad}\\
0 & 0 & -\frac{\cos(\theta)\sin(\phi)}{r} & -\frac{\cos(\phi)}{r}\\
0 & \frac{\cos(\theta)\sin(\phi)}{r} & 0 & -\frac{\sqrt{1-\kk r^2}}{r}\cot(\theta)\cos(\phi)\\
0 & \frac{\cos(\phi)}{r} & \frac{\sqrt{1-\kk r^2}}{r}\cot(\theta)\cos(\phi) & 0 
\end{bmatrix},\\
[\lambda_{X_5}]^a_{~b} &=\begin{bmatrix}
\phantom{\quad}0\phantom{\quad} & \phantom{\quad}0\phantom{\quad} & \phantom{\quad}0\phantom{\quad} & \phantom{\quad}0\phantom{\quad}\\
0 & 0 &  -\frac{\cos(\theta)\cos(\phi)}{r} & \frac{\sin(\phi)}{r}\\
0 & \frac{\cos(\theta)\cos(\phi)}{r} & 0 & \frac{\sqrt{1-\kk r^2}}{r}\cot(\theta)\sin(\phi)\\
0 & -\frac{\sin(\phi)}{r} & -\frac{\sqrt{1-\kk r^2}}{r}\cot(\theta)\sin(\phi) & 0 
\end{bmatrix},\\
[\lambda_{X_6}]^a_{~b} &=\begin{bmatrix}
\phantom{\quad}0\phantom{\quad} & \phantom{\quad}0\phantom{\quad} & \phantom{\quad}0\phantom{\quad} & \phantom{\quad}0\phantom{\quad}\\
0 & 0 &  \frac{\sin(\theta)}{r} & 0 \\
0 & -\frac{\sin(\theta)}{r} & 0 & 0\\
0 & 0 & 0 & 0 
\end{bmatrix}.
\end{align} 
\end{subequations}

With this information it becomes straightforward to determine the corresponding spin connection via equation \eqref{Lie spin connection}. 
Solving equation \eqref{Lie spin connection} we find the most general spin connection for Riemann-Cartan geometries assuming the diagonal orthonormal gauge respecting the $G_6$ group of motions on three dimensional spatial hypersurfaces to be

\begin{subequations} \label{RW_connection}
\text{\bf Riemann-Cartan} 
\begin{align}
\omega^1_{~2}&=\frac{W_1(t)}{\sqrt{1-\kk r^2}}\,dr \\
\omega^1_{~3}&=W_1(t)r\,d\theta \\
\omega^1_{~4}&=W_1(t)r\sin(\theta)d\phi \\
\omega^2_{~3}&=-\sqrt{1-\kk r^2}\,d\theta - W_2(t)r\sin(\theta)\,d\phi \\
\omega^2_{~4}&= W_2(t)r\,d\theta  -\sqrt{1- \kk r^2}\sin(\theta)\,d\phi \\
\omega^3_{~4}&=-\frac{W_2(t)}{\sqrt{1-\kk r^2}}\,dr-\cos(\theta)\,d\phi
\end{align}
\end{subequations}
where $W_j(t),\  j\in\{1,2\}$ are two arbitrary functions. This solution is equivalent to the frame spin connection pair determined in \cite{Coley:2022qug,McNutt:2023nxm}.

Additionally, restricting ourselves to the metric-teleparallel class of geometries, in which the curvature of the connection is zero, we obtain the most general spin connection for metric teleparallel geometries assuming the diagonal orthonormal gauge having a $G_6$ group of symmetries acting on three dimensional spatial surfaces.  Here we observe that there are three different solutions depending on the discrete parameter $\kk$ where both arbitrary functions become constants in equation \eqref{RW_connection}:

\begin{subequations} \label{TRW-connection}
\text{\bf Metric Teleparallel} 
\begin{align}
\kk &=1: &\ \  W_1(t)&=0,& W_2(t)&=\pm 1, \\
\kk &=0: &\ \  W_1(t)&=0,& W_2(t)&=0,\\
\kk &=-1:&\ \ W_1(t)&=\pm 1,& W_2(t)&=0.
\end{align}
\end{subequations}
The proper frame for these metric teleparallel geometries in which the spin connection is identically zero has a non-diagonal co-frame and can be found in \cite{Coley:2022qug} and as a special case in \cite{vandenHoogen:2024ash}.

\subsubsection{\texorpdfstring{$G_7$}{G7} --- Static Open and Closed Robertson-Walker}

The static open $(k=-1)$ and closed $(k=1)$ Robertson-Walker geometries are a simple extension of Section \ref{RW} with $\kk=\pm 1$ in which, in addition to the six Killing vectors acting on a 3-D spatial hyper-surface, there exists a seventh Killing vector $X_7=\partial_t$. The corresponding compensating matrix $[\lambda_{X_7}]^a_{~b}$ is identically zero.  Having this seventh Killing vector does not fundamentally change the nature of the corresponding co-frame and spin connection but restricts the functions 

\text{\bf Riemann-Cartan} 
\begin{equation}
A(t)=A_0,\qquad W_1(t)=W_1,\qquad W_2(t)=W_2,
\end{equation}
in equations \eqref{TRW-frame} and \eqref{RW_connection}.  For metric teleparallel geometries, one of $W_1(t)$ or $W_2(t)$ must be zero while the other is $\pm 1$ depending on the value of $\kk$ 

\begin{subequations} \label{TES-connection}
\text{\bf Metric Teleparallel} 
\begin{align}
\kk &=1: &\ \  W_1(t)&=0,& W_2(t)&=\pm 1, \\
\kk &=-1:&\ \ W_1(t)&=\pm 1,& W_2(t)&=0.
\end{align}
\end{subequations}
which is the same as equation \eqref{TRW-connection} for $\kk =\pm 1$. This metric teleparallel geometry with $k=1$ has been labelled the Teleparallel Einstein Static (TES) geometry in \cite{Coley:2022qug}.

\subsubsection{\texorpdfstring{$G_7$}{G7} --- Cosmological de Sitter}

The cosmological de Sitter geometry is a simple extension of Section \ref{RW} with $\kk=0$ in which, in addition to the six Killing vectors acting on a 3-dimensional spatial hyper-surface, we assume a seventh Killing vector $X_7=\partial_t-rH_0\partial_r$\footnote{We note that there are ten Killing vectors in general for a de Sitter geometry, but the existence of this seventh Killing vector with $\kk=0$ is sufficient to fix the metric function $A(t)=A_0e^{H_0t}$ to give us the cosmological version of the de Sitter geometry where $H_0$ is a constant Hubble parameter}. The corresponding compensating matrix $[\lambda_{X_7}]^a_{~b}$ is identically zero. For Riemann-Cartan geometries, having this seventh Killing vector does not fundamentally change the nature of the corresponding co-frame and spin connection but does restrict the arbitrary functions

\text{\bf Riemann-Cartan} 
\begin{equation}
A(t)=A_0e^{H_0t},\qquad W_1(t)=W_1e^{H_0t},\qquad W_2(t)=W_2e^{H_0t},
\end{equation}
in equations \eqref{TRW-frame} and \eqref{RW_connection}  where $A_0$, $W_1$ and $W_2$ are now constants.  
For metric teleparallel geometries, the spin connection is further restricted in that 

\text{\bf Metric-teleparallel}
\begin{equation}
W_1=W_2=0.
\end{equation}
This metric teleparallel version of de Sitter has been named Teleparallel de-Sitter (TDS) in \cite{Coley:2023dbg}. Below, we have a closer look at de Sitter and anti-de Sitter geometries in Riemann-Cartan and metric teleparallel geometries. We emphasize that in this section, the spin connection is invariant only under the seven dimensional subgroup, and not the full ten dimensional group of isometries of the metric.

\subsection{\texorpdfstring{$G_{10}$}{G10} --- de Sitter}

It is well known that a 4-dimensional geometry is maximally symmetric if it has a ten dimensional group of motions acting on the space. Assuming spherical coordinates $[t,r,\theta,\phi]$, the line-element can be expressed without loss of generality as
\begin{equation}\label{metric_DS}
ds^2 = -\left(1-kr^2 \right)dt^2 + \frac{1}{1-kr^2}\,dr^2+r^2\,d\theta^2+r^2\sin^2(\theta)\,d\phi^2,
\end{equation} 
where the discrete parameter $k$ can take on values $-1,0,1$ distinguishing the subcases known as Anti-de Sitter, Minkowski and de Sitter respectively. The analysis depends non-trivially on the parameter $k$, so in this subsection, we shall fix $k=1$, the de Sitter case. The anti-de Sitter case is considered on its own in the Section \ref{SectionADS}.

The ten vector generators of the symmetries are
\begin{subequations} \label{DS-generators}
\begin{align}
X_1&=\sin(\phi) \partial_\theta  +\cos(\phi) \cot(\theta) \partial_\phi, \\
X_2&=-\cos(\phi) \partial_\theta  +\sin(\phi) \cot(\theta) \partial_\phi, \\
X_3&=\partial_\phi, \\
X_4&=\partial_t,\\
X_5&= \frac{\cos(\theta)  e^{t}r }{\sqrt{1-r^2}}\partial_t
     +\sqrt{1-r^2}  \cos(\theta) e^{t} \partial_r 
     -\frac{\sqrt{1-r^2}  \sin(\theta) e^{t} }{r}\partial_\theta
, \\
X_6&= \frac{\sin(\theta) \sin(\phi) e^{t} r }{\sqrt{1-r^2}}\partial_t
      +\sqrt{1-r^2}  \sin(\theta)\sin(\phi) e^{t} \partial_r \nonumber \\ 
& \qquad \qquad \qquad 
      +\frac{\sqrt{1-r^2}  \cos(\theta) \sin(\phi) e^{t}  }{r}\partial_\theta
      +\frac{\sqrt{1-r^2}  \cos(\phi) e^{t}  }{r \sin(\theta)}\partial_\phi
, \\
X_7&= \frac{\sin(\theta) \cos(\phi)e^{t} r }{\sqrt{1-r^2}}\partial_t
      +\sqrt{1-r^2}  \sin(\theta)\cos(\phi) e^{t} \partial_r \nonumber \\ 
& \qquad \qquad \qquad 
     +\frac{\sqrt{1-r^2}\cos(\theta) \cos(\phi)  e^{t}  }{r}\partial_\theta
     -\frac{\sqrt{1-r^2}  \sin(\phi) e^{t} }{r \sin(\theta)}\partial_\phi
, \\
X_8&=\frac{\cos(\theta) e^{-t} r }{\sqrt{1-r^2}}\partial_t
    -\sqrt{1-r^2}  \cos(\theta) e^{-t} \partial_r 
    +\frac{\sqrt{1-r^2}  \sin(\theta) e^{-t} }{r}\partial_\theta
, \\
X_9&=\frac{\sin(\theta) \sin(\phi) e^{-t} r }{\sqrt{1-r^2}}\partial_t
    -\sqrt{1-r^2}  \sin(\theta)\sin(\phi) e^{-t} \partial_r \nonumber \\  
&\qquad \qquad \qquad 
    -\frac{\sqrt{1-r^2}\cos(\theta)  \sin(\phi) e^{-t} }{r}\partial_\theta
    -\frac{\sqrt{1-r^2}  \cos(\phi) e^{-t} }{r \sin(\theta)}\partial_\phi 
, \\
X_{10}&=\frac{\sin(\theta) \cos(\phi) e^{-t}r }{\sqrt{1-r^2}} \partial_t
        -\sqrt{1-r^2}  \sin(\theta)\cos(\phi) e^{-t} \partial_r \nonumber \\ 
&\qquad \qquad \qquad 
     -\frac{\sqrt{1-r^2}\cos(\theta) \cos(\phi)  e^{-t} }{r}\partial_\theta 
     +\frac{\sqrt{1-r^2} \sin(\phi) e^{-t} }{r \sin(\theta)}\partial_\phi ,
\end{align}
\end{subequations}

Choosing a diagonal orthonormal gauge for the co-frame basis we have
\begin{equation} \label{DS-frame}\renewcommand{\arraystretch}{1.4}
h^a=\begin{bmatrix}
  \sqrt{1-r^2}dt \\
\frac{1}{\sqrt{1-r^2}} \,dr \\
r \,d\theta \\
r \sin(\theta)\,d\phi \end{bmatrix}.
\end{equation}
See Section \ref{compDS} which shows the ten compensating matrices $[\lambda_{X_i}]^a_{~b}$.  With these matrices computed we can then determine the corresponding spin connection via equation \eqref{Lie spin connection}. Using the co-frame basis \eqref{DS-frame}, the most general spin connection in Riemann-Cartan geometries invariant under the full ten dimensional de Sitter group is of the form
 
\begin{subequations} \label{DSFULL_connection}
\text{\bf Riemann-Cartan} 
\begin{align}
\omega^1_{~2}&=-r\,dt, \\
\omega^2_{~3}&=-\sqrt{1-r^2}\,d\theta, \\
\omega^2_{~4}&=-\sqrt{1-r^2}\sin(\theta)\,d\phi, \\
\omega^3_{~4}&=-\cos(\theta)\,d\phi,
\end{align}
\end{subequations}
where all other components are zero.

Additionally, restricting ourselves to the metric-teleparallel class of geometries, in which the curvature of the connection is zero, we find that there is no solution. This stems from the fact that any 4-dimensional Riemann-Cartan geometry with non-trivial torsion admits at most a seven dimensional affine symmetry group \cite{Coley:2019zld}. It is however, possible to find a metric teleparallel solution to equation \eqref{Lie spin connection} if one restricts themselves to a seven dimensional subgroup of the full ten dimensional group of symmetries (see also \cite{Coley:2024tqe}). Indeed, there has been some questions in regard to the assumption of whether or not the metric and connection must be invariant under the same group of symmetries \cite{Bahamonde:2020snl}, we shall explore such an option below.

\subsubsection{The \texorpdfstring{$G_{7}$}{G7} subgroup of the \texorpdfstring{$G_{10}$}{G10} de Sitter group}\label{TDS}

As is already known, there is no co-frame/spin connection pair that satisfies the conditions for invariance under the full ten dimensional de Sitter and anti-de Sitter symmetry groups in metric teleparallel geometries unless the torsion is trivial, that is, the geometry is Minkowski. However the large ten dimensional group of isometries contains a seven dimensional subgroup, (the largest possible).  It has been suggested in the past that one defines the teleparallel de Sitter (TDS) as the metric teleparallel geometry that is invariant under this smaller group \cite{Coley:2022qug}.   

We define the Teleparallel de Sitter (TDS) geometries as those metric teleparallel geometries in which the metric is invariant under the ten dimensional de Sitter group but the spin connection is invariant only under a seven dimensional sub-group.  For the de Sitter metric in a metric teleparallel geometry, the corresponding seven dimensional subgroup has symmetry generating vectors 
\begin{equation}
\{X_1, X_2, X_3, X_4, X_5, X_6, X_7\}
\end{equation}
from equation \eqref{DS-generators}. The corresponding Lie algebra is
\begin{align}
[X_1,X_2]&=-X_3,        & [X_1,X_3]&= X_2,        & [X_1,X_5]=&-X_6,\nonumber \\
[X_1,X_6]&=X_5,         & [X_2,X_3]&=-X_1,        & [X_2,X_5]=&X_7,\nonumber\\
[X_2,X_7]&=-X_5,        & [X_3,X_6]&=X_7,         & [X_3,X_7]=&-X_6, \\
[X_4,X_5]&=X_5,         & [X_4,X_6]&=X_6,        & [X_4,X_7]=&X_7.\nonumber
\end{align}  
The $[\lambda_{X_i}]^a_{~b}$, $i\in\{1,2,3,4,5,6,7\}$ matrices are found in Section \ref{compDS}. It is now possible to solve equation \eqref{Lie spin connection} to yield the most general spin connection for Riemann-Cartan geometries invariant under the first $G_{7}$ subgroup of the de Sitter group.  Assuming the diagonal orthonormal gauge \eqref{ADS-frame} we find

\text{\bf Riemann-Cartan} 
\begin{subequations} \label{DS_connection_RC}
\begin{align}
\omega^1_{~2}&= W_1r\,dt - \frac{W_1+1}{1-r^2}\,dr,  \\
\omega^1_{~3}&=-\frac{(W_1+1)r}{\sqrt{1-r^2}}\,d\theta - \frac{W_2r^2}{\sqrt{1-r^2}}\sin(\theta)\,d\phi,  \\
\omega^1_{~4}&=\frac{W_2r^2}{\sqrt{1-r^2}}\,d\theta - \frac{(W_1+1)r}{\sqrt{1-r^2}}\sin(\theta)\,d\phi,    \\
\omega^2_{~3}&=-\frac{W_1r^2+1}{\sqrt{1-r^2}}\,d\theta-\frac{W_2r}{\sqrt{1-r^2}}\sin(\theta)\,d\phi,  \\
\omega^2_{~4}&=\frac{W_2r}{\sqrt{1-r^2}}\,d\theta - \frac{W_1r^2+1}{\sqrt{1-r^2}}\sin(\theta)\,d\phi, \\
\omega^3_{~4}&=W_2r\,dt-\frac{W_2}{1-r^2}\,dr-\cos(\theta)\,d\phi,
\end{align}
\end{subequations}
where $W_1$ and $W_2$ are arbitrary constants. The structure of the spin connection if one uses the other $G_7$ subgroup 
\begin{equation}
\{X_1, X_2, X_3, X_4, X_8, X_9, X_{10}\}
\end{equation}
is similar, only some of the signs change in the expression for the connection.

Additionally, if we restrict ourselves to the metric-teleparallel class of geometries, in which the curvature of the connection is zero, then the substitution

\text{\bf Metric Teleparallel} 
\begin{equation}\label{DS-connection_MT}
 W_1 =0, \qquad  W_2= 0,
\end{equation}
into equation \eqref{DS_connection_RC} yields the most general spin connection for metric teleparallel geometries assuming the diagonal orthonormal gauge \eqref{DS-frame} having the first $G_{7}$ subgroup of the de Sitter group.  This metric teleparallel geometry is the Teleparallel de Sitter geometry (TDS).

\subsection{\texorpdfstring{$G_{10}$}{G10} --- Anti-de Sitter}\label{SectionADS}

In this section we choose coordinates $[t,r,\theta,\phi]$ and express the anti-de Sitter metric as
\begin{equation}\label{metricADS}
ds^2=-\frac{1}{1-t^2} dt^2 +(1-t^2)dr^2+t^2d\theta^2+t^2\sinh^2(\theta)d\phi^2
\end{equation}
(This can be obtained from the de Sitter metric \eqref{metric_DS} via $t\to ir$, $r\to it$ and $\theta\to i\theta$ and $k=-1$.) 
Choosing a diagonal orthonormal gauge for the co-frame basis we have
\begin{equation} \label{ADS-frame}\renewcommand{\arraystretch}{1.4}
h^a=\begin{bmatrix}
  \frac{1}{\sqrt{1-t^2}}\,dt \\
    \sqrt{1-t^2} \,dr \\
t \,d\theta \\
t \sinh(\theta)\,d\phi \end{bmatrix}.
\end{equation}
The ten Killing vectors for the metric \eqref{metricADS} can be expressed as
\begin{subequations} \label{ADS-generators}
\begin{align}
X_1&=\sin(\phi) \partial_\theta  +\cos(\phi) \coth(\theta) \partial_\phi, \\
X_2&=-\cos(\phi) \partial_\theta  +\sin(\phi) \coth(\theta) \partial_\phi, \\
X_3&=\partial_\phi, \\
X_4&=\partial_r,\\
X_5&= \sqrt{1-t^2}\cosh(\theta)e^r \partial_t + \frac{t}{\sqrt{1-t^2}}\cosh(\theta)e^r \partial_r 
     -\frac{\sqrt{1-t^2}}{t}\sinh(\theta)e^r \partial_\theta, \\
X_6&=\sqrt{1-t^2}\sinh(\theta)\sin(\phi)e^r \partial_t +\frac{t}{\sqrt{1-t^2}}\sinh(\theta)\sin(\phi)e^r \partial_r 
     \nonumber\\ &\qquad \qquad\qquad-\frac{\sqrt{1-t^2}}{t}\cosh(\theta)\sin(\phi)e^r \partial_\theta 
     - \frac{\cos(\phi)}{\sinh(\theta)}\frac{\sqrt{1-t^2}}{t}e^r\partial_\phi, \\
X_7&=\sqrt{1-t^2}\sinh(\theta)\cos(\phi)e^r \partial_t +\frac{t}{\sqrt{1-t^2}}\sinh(\theta)\cos(\phi)e^r \partial_r 
     \nonumber\\ &\qquad \qquad\qquad-\frac{\sqrt{1-t^2}}{t}\cosh(\theta)\cos(\phi)e^r \partial_\theta 
     + \frac{\sin(\phi)}{\sinh(\theta)}\frac{\sqrt{1-t^2}}{t}e^r\partial_\phi, \\
X_8&= \sqrt{1-t^2}\cosh(\theta)e^{-r} \partial_t - \frac{t}{\sqrt{1-t^2}}\cosh(\theta)e^{-r} \partial_r 
     -\frac{\sqrt{1-t^2}}{t}\sinh(\theta)e^{-r} \partial_\theta, \\
X_9&=\sqrt{1-t^2}\sinh(\theta)\sin(\phi)e^{-r} \partial_t -\frac{t}{\sqrt{1-t^2}}\sinh(\theta)\sin(\phi)e^{-r} \partial_r 
     \nonumber\\ &\qquad \qquad\qquad-\frac{\sqrt{1-t^2}}{t}\cosh(\theta)\sin(\phi)e^{-r} \partial_\theta 
     - \frac{\cos(\phi)}{\sinh(\theta)}\frac{\sqrt{1-t^2}}{t}e^{-r}\partial_\phi, \\
X_{10}&=\sqrt{1-t^2}\sinh(\theta)\cos(\phi)e^{-r} \partial_t -\frac{t}{\sqrt{1-t^2}}\sinh(\theta)\cos(\phi)e^{-r} \partial_r 
     \nonumber\\ &\qquad \qquad\qquad-\frac{\sqrt{1-t^2}}{t}\cosh(\theta)\cos(\phi)e^{-r} \partial_\theta 
     + \frac{\sin(\phi)}{\sinh(\theta)}\frac{\sqrt{1-t^2}}{t}e^{-r}\partial_\phi, 
\end{align}
\end{subequations}
See Section \ref{comp_ADS} which show the ten compensating matrices $[\lambda_{X_i}]^a_{~b}$.  With these matrices computed we can then determine the corresponding spin connection via equation \eqref{Lie spin connection}. Using the co-frame basis \eqref{ADS-frame}, the most general spin connection in Riemann-Cartan geometries invariant under the full ten dimensional anti-de Sitter group is of the form

\text{\bf Riemann-Cartan} 
\begin{subequations} \label{ADS_connection_general}
\begin{align}
\omega^1_{~2}&=-t\,dr,  \\
\omega^1_{~3}&= \sqrt{1-t^2}\,d\theta,  \\
\omega^1_{~4}&=\sqrt{1-t^2}\sinh(\theta)\,d\phi,    \\
\omega^3_{~4}&=-\cosh(\theta)\,d\phi,
\end{align}
\end{subequations}
where all other components are zero.

If one attempts to determine the condition under which the above spin connection for Riemann-Cartan geometries has zero curvature, one finds that there is no solution. However, as before in the de Sitter case, if we assume that the spin connection is invariant under a smaller seven dimensional subgroup of the anti-de Sitter group, then it is possible to find a metric-teleparallel geometry that is invariant under this smaller subgroup.

\subsubsection{The \texorpdfstring{$G_{7}$}{G7} subgroup of the \texorpdfstring{$G_{10}$}{G10} anti-de Sitter group}\label{TADS}

The seven dimensional subgroup generated by $\{X_1,X_2,X_3,X_4,X_5,X_6,X_7\}$  has as its Lie algebra
\begin{align}
[X_1,X_2]&=X_3,         & [X_1,X_3]&=X_2,         & [X_1,X_5]=&X_6,\nonumber\\
[X_1,X_6]&=X_5,         & [X_2,X_3]&=-X_1,        & [X_2,X_5]=&-X_7,\nonumber\\
[X_2,X_7]&=-X_5,        & [X_3,X_6]&=X_7,         & [X_3,X_7]=&-X_6, \\
[X_4,X_5]&=X_5,         & [X_4,X_6]&=X_6,         & [X_4,X_7]=&X_7.\nonumber
\end{align}
A second seven dimensional subgroup exists with generators $\{X_1,X_2,X_3,X_4,X_8,X_9,X_{10}\}$. 
The $[\lambda_{X_i}]^a_{~b}$, $i\in\{1,2,3,4,5,6,7\}$ matrices are found in Section \ref{comp_ADS}. It is now possible to solve equation \eqref{Lie spin connection} to yield the most general spin connection for Riemann-Cartan geometries invariant under the first $G_{7}$ subgroup of the anti-de Sitter group.  Assuming the diagonal orthonormal gauge \eqref{ADS-frame} we find

\text{\bf Riemann-Cartan} 
\begin{subequations} \label{ADS_connection_RC}
\begin{align}
\omega^1_{~2}&=-\frac{W_1}{1-t^2}\,dt + (W_1-1)t\,dr,  \\
\omega^1_{~3}&=\frac{1-t^2+W_1t^2}{\sqrt{1-t^2}}\,d\theta + \frac{W_2t}{\sqrt{1-t^2}}\sinh(\theta)\,d\phi,  \\
\omega^1_{~4}&=-\frac{W_2t}{\sqrt{1-t^2}}\,d\theta + \frac{1-t^2+W_1t^2}{\sqrt{1-t^2}}\sinh(\theta)\,d\phi,    \\
\omega^2_{~3}&=\frac{W_1t}{\sqrt{1-t^2}}\,d\theta+\frac{W_2t^2}{\sqrt{1-t^2}}\sinh(\theta)\,d\phi,  \\
\omega^2_{~4}&=-\frac{W_2t^2}{\sqrt{1-t^2}}\,d\theta + \frac{W_1t}{\sqrt{1-t^2}}\sinh(\theta)\,d\phi, \\
\omega^3_{~4}&=-\frac{W_2}{1-t^2}\,dt+W_2t\,dr-\cosh(\theta)\,d\phi,
\end{align}
\end{subequations}
where $W_1$ and $W_2$ are arbitrary constants. The structure of the spin connection if one uses the other $G_7$ subgroup is similar, only some of the signs change.

Additionally, if we restrict ourselves to the metric-teleparallel class of geometries, in which the curvature of the connection is zero, then the substitution

\text{\bf Metric Teleparallel} 
\begin{equation}\label{ADS-connection_MT}
 W_1 =1, \qquad  W_2= 0,
\end{equation}
into equation \eqref{ADS_connection_RC} yields the most general spin connection for metric teleparallel geometries assuming the diagonal orthonormal gauge \eqref{ADS-frame} having the first $G_{7}$ subgroup of the anti-de Sitter group.  One can label this metric teleparallel geometry, the Teleparallel Anti-de Sitter geometry (TADS).


\section{Concluding remarks}\label{conclusion}

\subsection{Some Dynamical Comments}

While the primary goal of this work is to develop an alternative kinematical framework that could be useful in the study of non-Riemannian theories of gravity, some comments on what this means for the dynamics in these theories may be in order.  Below, we restrict ourselves to the class of $f(T)$ metric teleparallel theories of gravity as our sandbox to make some broad observations.

Consider the general co-frame \eqref{SSdiagframe} and zero curvature spin connection \eqref{orthonormal_connection_flat} compatible with spherical symmetry. This geometrical ansatz for spherical symmetry within $f(T)$ teleparallel gravity has been analyzed previously in \cite{Coley:2024tqe}.  Importantly, the anti-symmetric part of the field equations in \cite{Coley:2024tqe} are not identically satisfied even when all the symmetry conditions are imposed on the geometry apriori.  The antisymmetric part of the field equations in this case place constraints on the non-metric functions $\psi(t,r)$ and $\chi(t,r)$ which are present only in the spin connection (or in the proper co-frame).  In other words, the antisymmetric part of the $f(T)$ field equations constrain the part of the proper frame that is not present in the metric.  Alternatively, as a counter-example, consider the cosmological co-frame \eqref{TRW-frame} and corresponding zero curvature spin connection \eqref{RW_connection} with constraints \eqref{TRW-connection}: in this case the antisymmetric part of the $f(T)$ teleparallel gravity field equations are indeed identically satisfied for $k\in\{-1,0,1\}$. In short, imposing the full suite of symmetry conditions on the geometry could, but does not in general, yield a ``good''  tetrad \cite{Tamanini_Boehmer2012}, a tetrad that satisfies the anti-symmetric part of the field equations identically.

One may believe that one only needs to consider the isometries of the metric and that the antisymmetric part of $f(T)$ metric teleparallel gravity field equations will dictate the nature of the spin connection and consequently its symmetries.  However, there may be a problem with this approach.  Consider a Minkowski geometry having rectangular coordinates $x^\mu=[t,x,y,z]$ with the usual Minkowski metric $g_{\mu\nu}=Diag[-1,1,1,1]$. Choose a proper co-frame basis of the form $h^a=[dt,\cos(\chi(x))dx+\sin(\chi(x))dy,-\sin(\chi(x))dx+\cos(\chi(x))dy,dz]$ which depends on a differentiable function $\chi(x)$.  This arbitrary function $\chi(x)$ is present in the proper co-frame and can be transformed away via a Lorentz transformation to a new co-frame and a non-trivial spin connection.  In this example the torsion scalar is identically zero, so the field equations (symmetric and anti-symmetric parts) for the class of $f(T)$ metric teleparallel theories of gravity where $f(0)=0$ are identically satisfied in vacuum.  However, since the torsion two-form is non-trivial, there is geometrical structure that is not determined by the $f(T)$ field equations in vacuum.  Therefore, assuming a Minkowski metric and assuming that the antisymmetric part of $f(T)$ metric teleparallel gravity field equations are identically satisfied cannot be considered as sufficient ansatzes to determine a Minkowski geometry. In short, imposing the symmetry conditions on the metric only and solving for the ``good''  tetrad \cite{Tamanini_Boehmer2012}, cannot in general determine the connection that is invariant under the same symmetries as the metric.

Now of course, one could impose the isometries of the metric only, solve the antisymmetric part of the $f(T)$ field equations and determine the families of spin connections compatible with these equations.  Some members of this family will be invariant under the same group of symmetries as the metric and some may be invariant under a smaller subgroup of symmetries. The investigation of these wider classes of geometries in which the isometries of the metric are not fully imposed on the spin connection will yield a plethora of alternative gravitational models.

\subsection{Overview and Outlook}
We have illustrated a promising procedure that extends the idea of Killing vectors of the metric used extensively in the study of Riemannian geometries to Riemann-Cartan geometries which are not characterized solely by the metric, but through a co-frame and a spin-connection.  If given a metric, then it is known that members of the co-frame basis may no longer be invariant under Lie transport, but undergo an infinitesimal Lorentz transformation. This observation therefore necessitates a modification to the usual algorithms used in determining the initial geometrical framework when symmetries are involved in Riemann-Cartan geometries.
 
Here we illustrated a straightforward generalization of the ordinary Lie derivative to non-Riemannian, but metric compatible geometries.  The Lie-Lorentz derivative is a gauge covariant derivative suitable for investigating symmetries in non-Riemannian geometries. Choose any orthonormal co-frame compatible with the given metric, and set the Lie-Lorentz derivative of the metric(1) and co-frame(2) with respect to a vector field $X$ to zero.   If we assume that $X$ is a Killing vector for the metric, these conditions (1) and (2) determine the value of the compensating matrix.  With this compensating matrix fully determined one computes the independent spin connection compatible with that particular gauge choice for the co-frame. 

In particular, we have highlighted how the gauge covariant Lie derivative, or Lie-Lorentz derivative, can be employed to determine the initial geometrical framework (co-frame and its associated spin connection) for a number of Riemann-Cartan geometries. Details for the spherical symmetric and plane symmetric geometries are given for both Riemann-Cartan and metric teleparallel geometries.  The corresponding proper frame for spherical symmetry is known already \cite{vandenHoogen:2024ash}, but the proper frame for the plane symmetric case is new. Static spherically symmetric and Kantowski-Sachs geometries are subcases of the more general spherically symmetric model and are mentioned for completeness.  As an example of spatially homogeneous but anisotropic metrics having a non-trivial isotropy subgroup, the LRS Bianchi III was selected as an example.  The co-frame/spin connection pair was determined for both the Riemann-Cartan and metric teleparallel geometries, with the proper frame given in the appendix.  Further examples include the G\"{o}del metric which has a five dimensional isometry group and the Robertson Walker metric which has a six dimensional isometry group. Completing the set of examples, we looked at the maximally symmetric de Sitter and anti-de Sitter metrics in Riemann-Cartan geometries and we determined the appropriate co-frame and spin connection pair. the We also defined a metric teleparallel de Sitter geometry  and a metric teleparallel anti-de Sitter geometry by restricting the group of affine symmetries of the spin connection to a seven dimensional subgroup of the ten dimensional group of isometries (see also \cite{Coley:2022qug}). Reducing the number affine symmetries to a subgroup of the isometries of the metric could yield additional insight into other geometries and requires further consideration.


While spherically symmetric geometries have been well studied in Poincar\'{e} gauge and metric teleparallel theories of gravity, other classes of geometries have not.  The procedure outlined here is a kinematic approach and shows promise for being a straightforward mechanism to determine the initial geometrical framework to be used in investigating the dynamics in Poincar\'{e} gauge and metric teleparallel theories of gravity.  This procedure has the advantage that one need not determine an invariantly defined co-frame before trying to determine the spin connection. Any orthonormal co-frame computed from the metric will work.  The investigation of geometries outside of the traditional spherically symmetric and/or cosmologically symmetric geometries within a Poincar\'{e} gauge or metric teleparallel theory of gravity can now be accomplished. This will be the subject of future work.  We note that in its current form the proposed procedure is restricted to metric-compatible geometries, therefore any adaption of these ideas to non-metric compatible geometries, for example symmetric teleparallel geometries, requires a modified approach and is the subject of future studies.


\begin{acknowledgments}
RvdH was supported by the Natural Sciences and Engineering Research Council of Canada and the StFX University Council on Research. RvdH would like to acknowledge Elisabeth Grant who provided some administrative support during the undertaking of this project. The authors would like to thank David McNutt for some detailed comments and correspondence.  HF was supported by the Natural Sciences and Engineering Research Council of Canada through a Undergraduate Student Research Award.
\end{acknowledgments}



\providecommand{\href}[2]{#2}\begingroup\raggedright\endgroup


\appendix

\section{Appendix A: Proper frame in metric teleparallel geometries} \label{AppendixA}

\subsection{Proper frames}

In metric teleparallel geometries, in a neighborhood of a point, it is always possible to locally transform a given non-trivial spin connection to zero via a local Lorentz transformation.  This particular choice of gauge is called the Weitzenb\"{o}ck gauge, and the corresponding co-frame is called the proper co-frame.  We note that a proper co-frame may not exist globally depending on the topology.  In the bulk of this paper, we chose to do all computations in a co-frame that was simply derived from the assumed metric and consequently had a non-trivial spin connection. However, in this appendix we shall compute the local Lorentz transformation that will yield the corresponding proper co-frame for a number of the examples given in the text.

Assume an orthonormal co-frame basis $h^a$ and its associated non-trivial spin connection $\omega^a_{~b}$ as determined in Section \ref{examples}. In a metric-teleparallel geometry the spin connection can be generated by some local Lorentz transformation, $\Lambda^a_{~b}$ via
\begin{equation} \label{DE_conn}
\omega^a_{~b} = (\Lambda^{-1})^a_{~c}  d \Lambda^c_{~b}.
\end{equation}
To determine the proper co-frame ansatz $\tilde{h}^a$ for metric teleparallel geometries requires one to find the 
Lorentz transformation $\Lambda^a_{~b}$ that satisfies equation \eqref{DE_conn}.  One can then transform the given 
orthonormal co-frame $h^a$ and non-trivial spin connection $\omega^a_{~b}$ to a proper orthonormal co-frame
 $\tilde{h}^a=\Lambda^a_{~b}h^a$ having a trivial spin connection.

Rewriting equation \eqref{DE_conn} as
\begin{equation}
d \Lambda^a_{~b} = \Lambda^a_{~c} \omega^c_{~b},\label{DE_for_lambda}
\end{equation}
yields a system of linear partial differential equations that can be solved for the components, $\Lambda^a_{~b}$.
The integrability condition of \eqref{DE_for_lambda} is the zero curvature requirement of the connection $\omega^a_{~b}$,
\begin{equation}
R^a_{~b}:=d\omega^a_{~b}+\omega^a_{~c}\wedge \omega^c_{~b} = 0,
\end{equation}
which is also the requirement for metric teleparallel geometries.  
We also note that any solution to \eqref{DE_for_lambda} will contain a number of arbitrary constants of integration,
but these can be left factored out of the matrix expression for the solution. 
Given that the spin connection transforms covariantly under constant (global) Lorentz transformations, we can set this matrix of constants 
to the identity matrix without loss of generality.

In addition to determining the Lorentz transformation $\Lambda^a_{~b}$ we will also illustrate how this 
Lorentz transformation is a product of different boosts and rotations.

\subsection{Boosts and rotations}

Notation wise, the local Lorentz transformation is a $4 \times 4$ matrix $\Lambda$ with entries $\Lambda^a_{~b}$ which are typically functions of the coordinates. Equation \eqref{DE_for_lambda} can be expressed in matrix form as
\begin{equation}\label{de}
\Lambda_{,\mu}=\Lambda \Omega_\mu.
\end{equation}
where the $4\times 4$  matrix $\Omega_\mu$ represents the components of the spin connection and has entries $
\omega^a_{~b\mu}$.  Given the components of the spin connection one-form $\omega^a_{~b}$, we are able to set up the 
64 partial differential equations for a given geometrical ansatz.

The following matrices are used to represent the boosts and rotations in the $SO(1,3)$ Lorentz group; they are of the form:
\begin{equation} \label{elementary-Lorentz}
\begin{array}{cc}
B_x(\eta) :=
\begin{bmatrix}
\cosh(\eta) & \sinh(\eta) & 0 & 0 \\
\sinh(\eta) & \cosh(\eta) & 0 & 0 \\
0 & 0 & 1 & 0 \\
0 & 0 & 0 & 1
\end{bmatrix},
&\qquad
R_x(\eta) :=
\begin{bmatrix}
1 & 0 & 0 & 0 \\
0 & 1 & 0 & 0 \\
0 & 0 & \cos(\eta) & \sin(\eta) \\
0 & 0 & -\sin(\eta) & \cos(\eta)
\end{bmatrix},
\\
\\
B_y(\eta) :=
\begin{bmatrix}
\cosh(\eta) & 0 & \sinh(\eta) & 0 \\
0 & 1 & 0 & 0 \\
\sinh(\eta) & 0 & \cosh(\eta) & 0 \\
0 & 0 & 0 & 1
\end{bmatrix},
&\qquad
R_y(\eta) :=
\begin{bmatrix}
1 & 0 & 0 & 0 \\
0 & \cos(\eta) & 0 & \sin(\eta) \\
0 & 0 & 1 & 0 \\
0 & -\sin(\eta) & 0 & \cos(\eta)
\end{bmatrix},
\\
\\
B_z(\eta) :=
\begin{bmatrix}
\cosh(\eta) & 0 & 0 & \sinh(\eta) \\
0 & 1 & 0 & 0 \\
0 & 0 & 1 & 0 \\
\sinh(\eta) & 0 & 0 & \cosh(\eta)
\end{bmatrix},
&\qquad
R_z(\eta) :=
\begin{bmatrix}
1 & 0 & 0 & 0 \\
0 & \cos(\eta) & \sin(\eta) & 0 \\
0 & -\sin(\eta) & \cos(\eta) & 0 \\
0 & 0 & 0 & 1
\end{bmatrix},
\end{array}
\end{equation}
where the subscript on the boost matrices $B$ indicate the direction of the boost, and the subscript on the rotations $R$ indicate the axis of rotation.

\subsection{Proper frame for metric teleparallel plane-symmetric geometries}

\subsubsection{Case 1} \label{appendix-proper-plane}

In the case of a metric teleparallel plane symmetric geometry, the first solution for the spin
connection in equation \eqref{connection-plane}, yields values for the $\Omega_\mu$ matrices
\begin{align}
\Omega_{t} &=  \begin{bmatrix}
0 
& -\frac{d}{dt}\ln(W(t,x)) 
& 0 
& 0 \\
-\frac{d}{dt}\ln(W(t,x)) 
& 0 
& 0 
& 0 \\
0
& 0 
& 0 
& \frac{d}{dt}\chi(t,x) \\
0 & 0 & -\frac{d}{dt}\chi(t,x) & 0
\end{bmatrix},
\\
\Omega_{x} &= 
\begin{bmatrix}
0 
& -\frac{d}{dx}\ln(W(t,x)) 
& 0 
& 0 \\
-\frac{d}{dx}\ln(W(t,x)) 
& 0 
& 0 
& 0 \\
0
& 0 
& 0 
& \frac{d}{dx}\chi(t,x) \\
0 & 0 & -\frac{d}{dx}\chi(t,x) & 0
\end{bmatrix},
\\
\Omega_{y} &= 
W(t,x)\begin{bmatrix}
0 & 0 & \cos(\chi(t,x)) & \sin(\chi(t,x)) \\
0 & 0  & \cos(\chi(t,x)) & \sin(\chi(t,x)) \\
\cos(\chi(t,x)) & -\cos(\chi(t,x))  & 0 & 0 \\
\sin(\chi(t,x)) & -\sin(\chi(t,x)) & 0 & 0
\end{bmatrix},
\\
\Omega_{z} &= 
W(t,x)\begin{bmatrix}
0 & 0 & -\sin(\chi(t,x)) & \cos(\chi(t,x)) \\
0 & 0 & -\sin(\chi(t,x)) & \cos(\chi(t,x)) \\
-\sin(\chi(t,x)) & \sin(\chi(t,x))  & 0 & 0 \\
\cos(\chi(t,x)) & -\cos(\chi(t,x)) & 0 & 0
\end{bmatrix}.
\end{align}
The system of partial differential equations \eqref{de} 
can be solved through solving a sequence of differential equations.  
We shall start by solving the differential equation $\Lambda_{,t}=\Lambda\,  \Omega_t$.  Since the differential equation is linear, the solution can be determined through matrix exponentiation  and becomes
\begin{equation}
\Lambda(t,x,y,z)=\bar\Lambda(x,y,z) \,\text{exp}\left(\int \Omega_t\,dt\right).
\end{equation} 
In the plane symmetric case currently under consideration
\begin{align}
\text{exp}\left(\int \Omega_t\,dt\right) &= \left(
\begin{bmatrix} 1 & 0 & 0 & 0 \\ 0 & 1 & 0 & 0 \\ 0 & 0& 1 & 0 \\ 0 & 0& 0 & 1 \end{bmatrix} +
\begin{bmatrix} 0 & -\ln(W(t,x)) & 0 & 0 \\ -\ln(W(t,x)) & 0 & 0 & 0 \\ 0 & 0& 0 & \chi(t,x) \\ 0 & 0& -\chi(t,x) & 0 \end{bmatrix}\right.\nonumber \\
+&
\frac{1}{2!}\begin{bmatrix} (\ln(W(t,x))^2  & 0 & 0 & 0 \\ 0 & (\ln(W(t,x))^2 & 0 & 0 \\ 0 & 0& -(\chi(t,x))^2 & 0 \\ 0 & 0& 0 & -(\chi(t,x))^2 \end{bmatrix} 
\\
+&
\frac{1}{3!}\left.\begin{bmatrix} 0 & -(\ln(W(t,x)))^3 & 0 & 0 \\ -(\ln(W(t,x)))^3 & 0 & 0 & 0 \\ 0 & 0& 0 & -(\chi(t,x))^3\\ 0 & 0& (\chi(t,x))^3 & 0 \end{bmatrix} + \dots \right),\nonumber
\end{align} 
where we are able to identify the four different infinite series present as trigonometric or hyperbolic trigonometric functions. The solution to $\Lambda_{,t}=\Lambda\Omega_t$ is therefore
\begin{equation}\label{sol1}
\Lambda(t,x,y,z)=\bar\Lambda(x,y,z)\,  \Pi(t,x),
\end{equation}
where
\begin{eqnarray}
\Pi(t,x)&=&\text{exp}\left(\int \Omega_t\,dt\right)\\
&=&
\begin{bmatrix} 
\cosh(\ln(W(t,x)) & -\sinh(\ln(W(t,x)) & 0 & 0 \\
-\sinh(\ln(W(t,x)) & \cosh(\ln(W(t,x)) & 0 & 0 \\
0 & 0 & \cos(\chi(t,x)) & \sin(\chi(t,x)) \\
0 & 0  & -\sin(\chi(t,x)) & \cos(\chi(t,x)) 
\end{bmatrix}.\nonumber
\end{eqnarray}

We now substitute this partial solution \eqref{sol1} into the second differential equation $\Lambda_{,x}=\Lambda\Omega_x$ and rearrange to obtain
\begin{equation}
\bar\Lambda(x,y,z)_{,x}=\bar\Lambda(x,y,z)\,  \left(\Pi(t,x)\Omega_x -\Pi(t,x)_{,x}\right)\,  \Pi^{-1}(t,x) \equiv 0,
\end{equation}
which shows that $\bar\Lambda(x,y,z) = \bar\Lambda(y,z)$. 

We repeat the process for the differential equation $\Lambda_{,y}=\Lambda\Omega_y$.  The differential equation that results after substitution of the partial solution \eqref{sol1} with $\bar\Lambda(x,y,z) = \bar\Lambda(y,z)$ and some rearrangement yields
\begin{equation}
\bar\Lambda(y,z)_{,y} = \bar\Lambda(y,z) \,  \overline{\Omega}_y , 
\end{equation}
where
\begin{equation}
\overline{\Omega}_y =\Pi(t,x)\,  \Omega_y\,  \Pi^{-1}(t,x) \\
        =\begin{bmatrix} 0 & 0 & 1 & 0 \\ 0 & 0 & 1 & 0 \\ 1 & -1 & 0 & 0 \\ 0 & 0 & 0 & 0
        \end{bmatrix}.
\end{equation}
As $\overline{\Omega}_y$ is a constant matrix, we find that
\begin{equation}
\bar\Lambda(y,z) = \bar \Lambda(z)\,   \text{exp}\left(\overline{\Omega}_y y\right),
\end{equation}
where 
\begin{equation}
\text{exp}\left(\overline{\Omega}_y y\right) = I_4 + \overline{\Omega}_y y + (\overline{\Omega}_y)^2 \frac{y^2}{2}
 =\begin{bmatrix} 
 1+\frac{y^2}{2} & -\frac{y^2}{2} & \quad y & \quad 0 \\
\frac{y^2}{2} & 1-\frac{y^2}{2} & \quad y & \quad0 \\
y & -y & \quad 1 & \quad 0 \\
0 & 0 &  \quad 0 & \quad 1
\end{bmatrix}.
\end{equation}

We repeat the process for $\Lambda_{,z}=\Lambda\Omega_z$.  The differential equation that results is
\begin{equation}
\bar\Lambda(z)_{,z} = \bar\Lambda(z)\,   \overline{\Omega}_z,
\end{equation}
where
\begin{equation}
\overline{\Omega}_z =\text{exp}(\overline{\Omega}_y y)\,  \Pi(t,x)\,  \Omega_z\,  \Pi^{-1}(t,x)\,  \text{exp}(-\overline{\Omega}_y y) \\
        =\begin{bmatrix} 0 & 0 & 0 & 1 \\ 0 & 0 & 0 & 1 \\ 0 & 0 & 0 & 0 \\ 1 & -1 & 0 & 0
        \end{bmatrix}.
\end{equation}
As $\overline{\Omega}_z$ is a constant matrix, we find that
\begin{equation}
\bar\Lambda(z) = \Lambda_0 \,\text{exp}(\overline{\Omega}_z z),
\end{equation}
where 
\begin{equation}
\text{exp}(\overline{\Omega}_z z) = I_4 + \overline{\Omega}_z z + (\overline{\Omega}_z)^2 \frac{z^2}{2}
 =\begin{bmatrix} 
 1+\frac{z^2}{2} & -\frac{z^2}{2} & 0 & \quad z  \\
\frac{z^2}{2} & 1-\frac{z^2}{2} & 0 & \quad z \\
0 & 0 &  \quad 1 & \quad 0 \\
z & -z & \quad 0 & \quad 1 
\end{bmatrix},
\end{equation}
where $\Lambda_0$ is a constant matrix.  

After solving the differential equations by matrix exponentiation, the proceeding matrix solution is obtained  
\begin{eqnarray} \label{Lambda_plane1}
&&\Lambda(t,x,y,z) =\Lambda_0 \, \text{exp}(\overline{\Omega}_z z) \, \text{exp}(\overline{\Omega}_y y)\, \Pi(t,x) \nonumber\\
\nonumber\\
&&\quad=  \Lambda_0 \scriptsize
\begin{bmatrix}
\frac{1+(y^2+z^2+1)W(t,x)^2}{2W(t,x)}
& \frac{1-(y^2+z^2+1)W(t,x)^2}{2W(t,x)}
& y\cos(\chi(t,x)) - z\sin(\chi(t,x))
& y\sin(\chi(t,x)) + z\cos(\chi(t,x)) \\
\frac{1+(y^2+z^2-1)W(t,x)^2}{2W(t,x)}
& \frac{1-(y^2+z^2-1)W(t,x)^2}{2W(t,x)}
& y\cos(\chi(t,x)) - z\sin(\chi(t,x))
& y\sin(\chi(t,x)) + z\cos(\chi(t,x)) \\
yW(t,x)
& -yW(t,x) 
& \cos{(\chi(t,x))}
& \sin{(\chi(t,x))} \\
zW(t,x) & -zW(t,x) & -\sin{(\chi(t,x))} & \cos{(\chi(t,x))}
\end{bmatrix}.\nonumber
\end{eqnarray}
\normalsize
The matrix $\Lambda_0$ can be set to the $4 \times 4$ identity matrix $I_4$ without loss of generality.  

The matrix $\Lambda(t,x,y,z)$, can be factored into elementary Lorentz matrices \eqref{elementary-Lorentz}, 
\begin{equation}
    \Lambda(t,x,y,z) =B_x(\ln z)\,  C_y\,  B_x(-\ln z + \ln y)\,  C_z\,  B_x(-\ln y - \ln W(t,x))\,  R_x(\chi(t,x)),
\end{equation}
where $C_y$, $C_z$ are global (i.e., constant) Lorentz matrices
\begin{subequations}
    \begin{align}
    C_y &= R_y(\tan^{-1}(2)) \, B_x\left(\tanh^{-1}\left(-\frac{1}{\sqrt{5}}\right)\right) \, B_z\left(\tanh^{-1}\left(\frac{2}{3}\right)\right),\\
    C_z &= R_z(\tan^{-1}(2)) \, B_x\left(\tanh^{-1}\left(-\frac{1}{\sqrt{5}}\right)\right)\, B_y\left(\tanh^{-1}\left(\frac{2}{3}\right)\right).
\end{align}
\end{subequations} 
Multiplying the matrix in equation \eqref{Lambda_plane1} by the co-frame in equation \eqref{Plane-frame} will yield a proper frame for metric teleparallel plane symmetric geometries. 

\subsubsection{Case 2} 

In the case of a metric teleparallel plane symmetric geometry, the second solution for the spin connection in
 equation \eqref{connection-plane2}, yields values for the $\Omega_\mu$ matrices
\begin{align}
\Omega_{t} = 
\begin{bmatrix}
0 & \frac{d\psi}{dt} & 0 & 0 \\
\frac{d\psi}{dt} & 0 & 0 & 0 \\
0 & 0 & 0 & \frac{d\chi}{dt} \\
0 & 0 & -\frac{d\chi}{dt} & 0
\end{bmatrix},
& \qquad 
\Omega_{x} = 
\begin{bmatrix}
0 & \frac{d\psi}{dx} & 0 & 0 \\
\frac{d\psi}{dx} & 0 & 0 & 0 \\
0 & 0 & 0 & \frac{d\chi}{dx} \\
0 & 0 & -\frac{d\chi}{dx} & 0
\end{bmatrix},
\end{align}
where $\Omega_{y}$ and $\Omega_z$ are zero matrices. We solve equation \eqref{DE_for_lambda} to find that the corresponding Lorentz transformation to be
\begin{equation} \label{Lambda-plane2}
\Lambda=\begin{bmatrix}
\cosh(\psi(t,x))              & \sinh(\psi(t,x))        & 0               & 0 \\
\sinh(\psi(t,x))              & \cosh(\psi(t,x))        & 0               & 0  \\
0                             & 0                      & \cos(\chi(t,x))  & \sin(\chi(t,x) \\
0                             & 0                      & -\sin(\chi(t,x)) & \cos(\chi(t,x))
\end{bmatrix}.
\end{equation}
The Lorentz transformation \eqref{Lambda-plane2} is easily seen to be the product of a local boost and a local rotation 
\begin{equation}
\Lambda(t,y) = B_x(\psi(t,x)) R_x(\chi(t,x)).
\end{equation}
Multiplying the matrix in equation \eqref{Lambda-plane2} by the co-frame in equation \eqref{Plane-frame} will yield a proper frame for metric teleparallel plane symmetric geometries. 

\subsection{Proper frame for metric-teleparallel LRS Bianchi III geometries}\label{appendix-proper-LRS-Bianchi}

In the case of a metric teleparallel LRS Bianchi type III geometry, the solution for the spin connection in
equation \eqref{solutionLRSIII-NULLCURV}, yields values for the $\Omega_\mu$ matrices
\begin{subequations}
\begin{align}
\Omega_{t} &= \begin{bmatrix}
0 & 0 & 0 & \frac{d\psi}{dt}  \\
0 & 0 & -\frac{d\chi}{dt}  & 0 \\
0 & \frac{d\chi}{dt}  & 0 & 0 \\
\frac{d\psi}{dt}  & 0 & 0 & 0
\end{bmatrix},
\\
\Omega_{x} &= \begin{bmatrix}
0 & \cos{(\chi)}\cosh{(\psi)} & -\sin{(\chi)}\cosh{(\psi)} & 0 \\
\cos{(\chi)}\cosh{(\psi)} & 0 & 0 & \cos{(\chi)}\sinh{(\psi)} \\
-\sin{(\chi)}\cosh{(\psi)} & 0 & 0 & -\sin{(\chi)}\sinh{(\psi)} \\
0 & -\cos{(\chi)}\sinh{(\psi)} & \sin{(\chi)}\sinh{(\psi)} & 0
\end{bmatrix},
\\
\Omega_{y} &= e^x\begin{bmatrix}
0 & \sin{(\chi)}\cosh{(\psi)} & \cos{(\chi)}\cosh{(\psi)} & 0 \\
\sin{(\chi)}\cosh{(\psi)} & 0 & -1 & \sin{(\chi)}\sinh{(\psi)}\\
\cos{(\chi)}\cosh{(\psi)} & 1 & 0 & \cos{(\chi)}\sinh{(\psi)} \\
0 & -\sin{(\chi)}\sinh{(\psi)} & -\cos{(\chi)}\sinh{(\psi)} & 0
\end{bmatrix},
\end{align}
\end{subequations}
where $\psi=\psi(t)$, $\chi=\chi(t)$ and
where the matrix $\Omega_{z}$ is the zero matrix. 

We can solve equation \eqref{DE_for_lambda} in the same manner as outlined in Section \ref{appendix-proper-plane} to find a lengthy expression for $\Lambda$ as a function of $x$, $y$, $\chi(t)$ and $\psi(t)$.  We can express 
the corresponding Lorentz transformation as the following product  of elementary Lorentz matrices \eqref{elementary-Lorentz}
\begin{equation}\label{Lambda-LRS}
\Lambda(t,x,y) 
= B_x(\alpha(y))
  C_z
  B_x(-\alpha(y)) 
  B_x(x) B_z(\psi(t)) R_z(-\chi(t)).
\end{equation}
where $\alpha(y)$ is given by
\begin{equation}
\alpha(y)=\tanh^{-1}\left(\frac{1-y^2}{1+y^2}\right),
\end{equation}
and $C_z$ is a constant matrix 
\begin{equation}
C_z=R_z\left(-\tan^{-1}(2)\right)B_x\left(\tanh^{-1}\left(\frac{1}{\sqrt{5}}\right)\right)B_y\left(\tanh^{-1}\left(\frac{2}{3}\right)\right).
\end{equation}
Multiplying the matrix in equation \eqref{Lambda-LRS} by the co-frame in equation \eqref{LRSIII-frame} will yield a proper frame for metric teleparallel LRS Bianchi type III geometries.

\subsection{Proper frame for metric teleparallel G\"{o}del geometries}\label{appendix-proper-godel}

\subsubsection{Case 1}

In the case of a metric teleparallel G\"{o}del geometry, the first solution for the spin connection in
 equation \eqref{Godel-spin-1}, yields values for the $\Omega_\mu$ matrices
\begin{align}
\Omega_{t} = 
\begin{bmatrix}
0 & 0 & 0 & 0 \\
0 & 0 & 0 & \frac{1}{\sqrt{2}} \\
0 & 0 & 0 & 0 \\
0 & -\frac{1}{\sqrt{2}} & 0 & 0
\end{bmatrix},
& \qquad 
\Omega_{y} = 
\begin{bmatrix}
0 & 0 & W_4 & 0 \\
0 & 0 & 0 & W_8 \\
W_4 & 0 & 0 & 0 \\
0 & -W_8 & 0 & 0
\end{bmatrix},
\end{align}
where $\Omega_{x}$ and $\Omega_z$ are zero matrices and $W_4$ and $W_8$ are constants.
We solve equation \eqref{DE_for_lambda} to find that the corresponding Lorentz transformation to be
\begin{equation} \label{Lambda-godel1}
\Lambda=\begin{bmatrix}
\cosh(W_4y)                      & 0                            & \sinh(W_4y)  & 0 \\
0                                 & \cos(W_8y+\frac{t}{\sqrt{2}}) & 0            & \sin(W_8y+\frac{t}{\sqrt{2}}) \\
\sinh(W_4y)                      & 0                            & \cosh(W_4y)  & 0 \\
0                                 &-\sin(W_8y+\frac{t}{\sqrt{2}}) & 0            & \cos(W_8y+\frac{t}{\sqrt{2}})
\end{bmatrix}.
\end{equation}
The Lorentz transformation \eqref{Lambda-godel1} is easily seen to be the product of a local boost and a local rotation 
\begin{equation}
\Lambda(t,y) = B_y(W_4 y) R_y(W_8 y+\frac{t}{\sqrt{2}}).
\end{equation}
Multiplying the matrix in equation \eqref{Lambda-godel1} by the co-frame in equation \eqref{Godel-frame} will yield a proper frame for metric teleparallel G\"{o}del geometries.

\subsubsection{Case 2}
In the case of a metric teleparallel G\"{o}del geometry, the second solution for the spin connection in
 equation \eqref{Godel-spin-2}, yields values for the $W_\mu$ matrices
\begin{subequations}
\begin{align}
\Omega_{x} &= 
\begin{bmatrix}
0 & \cosh(\psi)\cos(\chi) & 0 & \cosh(\psi)\sin(\chi) \\
\cosh(\psi)\cos(\chi) & 0 & \sinh(\psi)\cos(\chi) & 0 \\
0 & -\sinh(\psi)\cos(\chi) & 0 & -\sinh(\psi)\sin(\chi) \\
\cosh(\psi)\sin(\chi) & 0 & \sinh(\psi)\sin(\chi) & 0
\end{bmatrix},\\
\Omega_{z} &= \frac{e^x}{\sqrt{2}}
\begin{bmatrix}
0 & -\cosh(\psi)\sin(\chi) & 0 & \cosh(\psi)\cos(\chi) \\
-\cosh(\psi)\sin(\chi) & 0 & -\sinh(\psi)\sin(\chi) & -1 \\
0 &  \sinh(\psi)\sin(\chi) & 0 & -\sinh(\psi)\cos(\chi) \\
\cosh(\psi)\cos(\chi) & 1 & \sinh(\psi)\cos(\chi) & 0
\end{bmatrix},
\end{align}
\end{subequations}
where $\Omega_t$ and $\Omega_y$ are zero matrices and $\chi$ and $\psi$ are constants.

We can solve equation \eqref{DE_for_lambda} in the same manner as outlined in Section \ref{appendix-proper-plane} to find a lengthy expression for $\Lambda$ as a function of $x$ and $z$.  However we can express 
the corresponding Lorentz transformation as the following 
product  of elementary Lorentz matrices \eqref{elementary-Lorentz}
\begin{equation}\label{Lambda-godel2}
\Lambda(x,z) 
= R_{y}(-\chi)B_y(-\psi)  
R_y(-\alpha(z)) B_z(\beta(z)) R_{y}(-\alpha(z))B_x(x)  
B_y(\psi) R_y(\chi).
\end{equation}
where $\chi$ and $\psi$ are constants and $\alpha(z)$ and $\beta(z)$ are determined by the following
\begin{align}
\alpha(z) &=\tan^{-1}\left(\frac{z}{2\sqrt{2}}\right),&
\beta(z) &=\tanh^{-1}\left(\frac{{z}{\sqrt{z^2+8}}}{z^2+4}\right).
\end{align}
Multiplying the matrix in equation \eqref{Lambda-godel2} by the co-frame in equation \eqref{Godel-frame} will yield a proper frame for metric teleparallel G\"{o}del geometries.

\section{Compensating Matrices}

\subsection{de Sitter}\label{compDS}
The compensating matrices $[\lambda_{X_i}]^a_{~b}$ for the de Sitter metric \eqref{metric_DS} with the orthonormal co-frame \eqref{DS-frame} are
\begin{subequations}
\renewcommand{\arraystretch}{1.2}
\begin{align}
[\lambda_{X_1}]^a_{~b} &=\begin{bmatrix}
\phantom{\qquad}0\phantom{\qquad} & \phantom{\qquad}0\phantom{\qquad} & \phantom{\qquad}0\phantom{\qquad} & \phantom{\qquad}0\phantom{\qquad}\\
0 & 0 & 0 & 0\\
0 & 0 & 0 & -\cos(\phi)\csc(\theta)\\
0 & 0 & \cos(\phi)\csc(\theta) & 0 
\end{bmatrix},
\\
[\lambda_{X_2}]^a_{~b} &=\begin{bmatrix}
\phantom{\qquad}0\phantom{\qquad} & \phantom{\qquad}0\phantom{\qquad} & \phantom{\qquad}0\phantom{\qquad} & \phantom{\qquad}0\phantom{\qquad}\\
0 & 0 & 0 & 0\\
0 & 0 & 0 & -\sin(\phi)\csc(\theta)\\
0 & 0 & \sin(\phi)\csc(\theta) & 0 
\end{bmatrix},\\
[\lambda_{X_5}]^a_{~b} &=\begin{bmatrix}
0                                                        & -\frac{\cos(\theta)e^{t}}{\sqrt{1-r^2}} &  \sin(\theta)e^{t} & 0\\
-\frac{\cos(\theta)e^{t}}{\sqrt{1-r^2}} & 0                                      & \frac{\sin(\theta)e^{t}}{r}           & 0\\
\sin(\theta)e^{t}                        & -\frac{\sin(\theta)e^{t}}{r}                       & 0                           & 0 \\
\phantom{\quad}0\phantom{\quad}                          & \phantom{\quad}0\phantom{\quad} & \phantom{\quad}0\phantom{\quad} & \phantom{\quad}0\phantom{\quad}
\end{bmatrix},\\
[\lambda_{X_6}]^a_{~b} &=\begin{bmatrix}
\phantom{\quad}0\phantom{\quad}                                    & -\frac{\sin(\theta)\sin(\phi)e^{t}}{\sqrt{1-r^2}} & -\cos(\theta)\sin(\phi)e^{t}   & -\cos(\phi)e^{t}\\
-\frac{\sin(\theta)\sin(\phi)e^{t}}{\sqrt{1-r^2}} & 0                                                                  &  -\frac{\cos(\theta)\sin(\phi)e^{t}}{r} & -\frac{\cos(\phi)e^{t}}{r}\\
-\cos(\theta)\sin(\phi)e^{t}                      &  \frac{\cos(\theta)\sin(\phi)e^{t}}{r}                     & 0                                               & -\frac{\sqrt{1-r^2}\cot(\theta)\cos(\phi)e^{t}}{r}\\
-\cos(\phi)e^{t}                                   & \frac{\cos(\phi)e^{t}}{r}                                  & \frac{\sqrt{1-r^2}\cot(\theta)\cos(\phi)e^{t}}{r} & 0 
\end{bmatrix},\\
[\lambda_{X_7}]^a_{~b} &=\begin{bmatrix}
\phantom{\quad}0\phantom{\quad}                                    & -\frac{\sin(\theta)\cos(\phi)e^{t}}{\sqrt{1-r^2}} & -\cos(\theta)\cos(\phi)e^{t}    & \sin(\phi)e^{t}\\
-\frac{\sin(\theta)\cos(\phi)e^{t}}{\sqrt{1-r^2}} & 0                                                                  &  -\frac{\cos(\theta)\cos(\phi)e^{t}}{r}  & \frac{\sin(\phi)e^{t}}{r}\\
-\cos(\theta)\cos(\phi)e^{t}                      &  \frac{\cos(\theta)\cos(\phi)e^{t}}{r}                     & 0                                                & \frac{\sqrt{1-r^2}\cot(\theta)\sin(\phi)e^{t}}{r}\\
\sin(\phi)e^{t}                                    & -\frac{\sin(\phi)e^{t}}{r}                                 & -\frac{\sqrt{1-r^2}\cot(\theta)\sin(\phi)e^{t}}{r} & 0 
\end{bmatrix}.\\
[\lambda_{X_8}]^a_{~b} &=\begin{bmatrix}
0                                                         & -\frac{\cos(\theta)e^{-t}}{\sqrt{1-r^2}}   & \sin(\theta)e^{-t} & 0\\
-\frac{\cos(\theta)e^{-t}}{\sqrt{1-r^2}} & 0                                                           & -\frac{\sin(\theta)e^{-t}}{r}           & 0\\
\sin(\theta)e^{-t}                        & \frac{\sin(\theta)e^{-t}}{r}                       & 0                           & 0 \\
\phantom{\quad}0\phantom{\quad}                           & \phantom{\quad}0\phantom{\quad} & \phantom{\quad}0\phantom{\quad} & \phantom{\quad}0\phantom{\quad}
\end{bmatrix},\\
[\lambda_{X_9}]^a_{~b} &=\begin{bmatrix}
\phantom{\quad}0\phantom{\quad}                                    & -\frac{\sin(\theta)\sin(\phi)e^{-t}}{\sqrt{1-r^2}} & -\cos(\theta)\sin(\phi)e^{-t}   & -\cos(\phi)e^{-t}\\
-\frac{\sin(\theta)\sin(\phi)e^{-t}}{\sqrt{1-r^2}} & 0                                                                  &  \frac{\cos(\theta)\sin(\phi)e^{-t}}{r} & \frac{\cos(\phi)e^{-t}}{r}\\
-\cos(\theta)\sin(\phi)e^{-t}                     &  -\frac{\cos(\theta)\sin(\phi)e^{-t}}{r}                     & 0                                               & \frac{\sqrt{1-r^2}\cot(\theta)\cos(\phi)e^{-t}}{r}\\
-\cos(\phi)e^{-t}                                   & -\frac{\cos(\phi)e^{-t}}{r}                                  & -\frac{\sqrt{1-r^2}\cot(\theta)\cos(\phi)e^{-t}}{r} & 0
\end{bmatrix},\\
[\lambda_{X_{10}}]^a_{~b} &=\begin{bmatrix}
\phantom{\quad}0\phantom{\quad}                                    & -\frac{\sin(\theta)\cos(\phi)e^{-t}}{\sqrt{1-r^2}} & -\cos(\theta)\cos(\phi)e^{-t}    & \sin(\phi)e^{-t}\\
-\frac{\sin(\theta)\cos(\phi)e^{-t}}{\sqrt{1-r^2}} & 0                                                                  &  \frac{\cos(\theta)\cos(\phi)e^{-t}}{r}  & -\frac{\sin(\phi)e^{-t}}{r}\\
-\cos(\theta)\cos(\phi)e^{-t}                      &  -\frac{\cos(\theta)\cos(\phi)e^{-t}}{r}                     & 0                                                & -\frac{\sqrt{1-r^2}\cot(\theta)\sin(\phi)e^{-t}}{r}\\
\sin(\phi)e^{-t}                                    & \frac{\sin(\phi)e^{-t}}{r}                                 & \frac{\sqrt{1-r^2}\cot(\theta)\sin(\phi)e^{-t}}{r} & 0 
\end{bmatrix},
\end{align} 
\end{subequations}
with $[\lambda_{X_3}]^a_{~b}$ and $[\lambda_{X_4}]^a_{~b}$ only having zero entries.

\subsection{Anti-de Sitter}\label{comp_ADS}

The compensating matrices $[\lambda_{X_i}]^a_{~b}$ for the anti-de Sitter metric \eqref{metricADS} with the orthonormal co-frame \eqref{ADS-frame} are
\begin{subequations}
\renewcommand{\arraystretch}{1.2}
\begin{align}
[\lambda_{X_1}]^a_{~b} &=\begin{bmatrix}
\phantom{\qquad}0\phantom{\qquad} & \phantom{\qquad}0\phantom{\qquad} & \phantom{\qquad}0\phantom{\qquad} & \phantom{\qquad}0\phantom{\qquad}\\
0 & 0 & 0 & 0\\
0 & 0 & 0 & -\cos(\phi)\csch(\theta)\\
0 & 0 & \cos(\phi)\csch(\theta) & 0 
\end{bmatrix},
\\
[\lambda_{X_2}]^a_{~b} &=\begin{bmatrix}
\phantom{\qquad}0\phantom{\qquad} & \phantom{\qquad}0\phantom{\qquad} & \phantom{\qquad}0\phantom{\qquad} & \phantom{\qquad}0\phantom{\qquad}\\
0 & 0 & 0 & 0\\
0 & 0 & 0 & -\sin(\phi)\csch(\theta)\\
0 & 0 & \sin(\phi)\csch(\theta) & 0 
\end{bmatrix},\\
[\lambda_{X_5}]^a_{~b} &=\begin{bmatrix}
0                                      & -\frac{\cosh(\theta)e^r}{\sqrt{1-t^2}} & -\frac{\sinh(\theta)e^r}{t} & 0\\
-\frac{\cosh(\theta)e^r}{\sqrt{1-t^2}} & 0                                      & -\sinh(\theta)e^r           & 0\\
-\frac{\sinh(\theta)e^r}{t}            & \sinh(\theta)e^r                       & 0                           & 0 \\
\phantom{\quad}0\phantom{\quad} & \phantom{\quad}0\phantom{\quad} & \phantom{\quad}0\phantom{\quad} & \phantom{\quad}0\phantom{\quad}
\end{bmatrix},\\
[\lambda_{X_6}]^a_{~b} &=\begin{bmatrix}
\phantom{\quad}0\phantom{\quad} & -\frac{\sinh(\theta)\sin(\phi)e^r}{\sqrt{1-t^2}} & -\frac{\cosh(\theta)\sin(\phi)e^r}{t} & -\frac{\cos(\phi)e^r}{t}\\
-\frac{\sinh(\theta)\sin(\phi)e^r}{\sqrt{1-t^2}} & 0 &  -\cosh(\theta)\sin(\phi)e^r & -\cos(\phi)e^r\\
-\frac{\cosh(\theta)\sin(\phi)e^r}{t} &  \cosh(\theta)\sin(\phi)e^r & 0 & \frac{\sqrt{1-t^2}\coth(\theta)\cos(\phi)e^r}{t}\\
-\frac{\cos(\phi)e^r}{t} & \cos(\phi)e^r & -\frac{\sqrt{1-t^2}\coth(\theta)\cos(\phi)e^r}{t} & 0 
\end{bmatrix},\\
[\lambda_{X_7}]^a_{~b} &=\begin{bmatrix}
\phantom{\quad}0\phantom{\quad} & -\frac{\sinh(\theta)\cos(\phi)e^r}{\sqrt{1-t^2}} & -\frac{\cosh(\theta)\cos(\phi)e^r}{t} & \frac{\sin(\phi)e^r}{t}\\
-\frac{\sinh(\theta)\cos(\phi)e^r}{\sqrt{1-t^2}} & 0 &  -\cosh(\theta)\cos(\phi)e^r & \sin(\phi)e^r\\
-\frac{\cosh(\theta)\cos(\phi)e^r}{t} &  \cosh(\theta)\cos(\phi)e^r & 0 & -\frac{\sqrt{1-t^2}\coth(\theta)\sin(\phi)e^r}{t}\\
\frac{\sin(\phi)e^r}{t} & -\sin(\phi)e^r & \frac{\sqrt{1-t^2}\coth(\theta)\sin(\phi)e^r}{t} & 0 
\end{bmatrix}.\\
[\lambda_{X_8}]^a_{~b} &=\begin{bmatrix}
0                                      & \frac{\cosh(\theta)e^{-r}}{\sqrt{1-t^2}} & -\frac{\sinh(\theta)e^{-r}}{t} & 0\\
\frac{\cosh(\theta)e^{-r}}{\sqrt{1-t^2}} & 0                                      & \sinh(\theta)e^{-r}           & 0\\
-\frac{\sinh(\theta)e^{-r}}{t}            & -\sinh(\theta)e^{-r}                       & 0                           & 0 \\
\phantom{\quad}0\phantom{\quad} & \phantom{\quad}0\phantom{\quad} & \phantom{\quad}0\phantom{\quad} & \phantom{\quad}0\phantom{\quad}
\end{bmatrix},\\
[\lambda_{X_9}]^a_{~b} &=\begin{bmatrix}
\phantom{\quad}0\phantom{\quad} & \frac{\sinh(\theta)\sin(\phi)e^{-r}}{\sqrt{1-t^2}} & -\frac{\cosh(\theta)\sin(\phi)e^{-r}}{t} & -\frac{\cos(\phi)e^{-r}}{t}\\
\frac{\sinh(\theta)\sin(\phi)e^{-r}}{\sqrt{1-t^2}} & 0 &  \cosh(\theta)\sin(\phi)e^{-r} & \cos(\phi)e^{-r}\\
-\frac{\cosh(\theta)\sin(\phi)e^{-r}}{t} &  -\cosh(\theta)\sin(\phi)e^{-r} & 0 & \frac{\sqrt{1-t^2}\coth(\theta)\cos(\phi)e^{-r}}{t}\\
-\frac{\cos(\phi)e^{-r}}{t} & -\cos(\phi)e^{-r} & -\frac{\sqrt{1-t^2}\coth(\theta)\cos(\phi)e^{-r}}{t} & 0 
\end{bmatrix},\\
[\lambda_{X_{10}}]^a_{~b} &=\begin{bmatrix}
\phantom{\quad}0\phantom{\quad} & \frac{\sinh(\theta)\cos(\phi)e^{-r}}{\sqrt{1-t^2}} & -\frac{\cosh(\theta)\cos(\phi)e^{-r}}{t} & \frac{\sin(\phi)e^{-r}}{t}\\
\frac{\sinh(\theta)\cos(\phi)e^{-r}}{\sqrt{1-t^2}} & 0 &  \cosh(\theta)\cos(\phi)e^{-r} & -\sin(\phi)e^{-r}\\
-\frac{\cosh(\theta)\cos(\phi)e^{-r}}{t} &  -\cosh(\theta)\cos(\phi)e^{-r} & 0 & -\frac{\sqrt{1-t^2}\coth(\theta)\sin(\phi)e^{-r}}{t}\\
\frac{\sin(\phi)e^{-r}}{t} & \sin(\phi)e^{-r} & \frac{\sqrt{1-t^2}\coth(\theta)\sin(\phi)e^{-r}}{t} & 0 
\end{bmatrix},
\end{align} 
\end{subequations}
with $[\lambda_{X_3}]^a_{~b}$ and $[\lambda_{X_4}]^a_{~b}$ only having zero entries.

\end{document}